\documentclass[secnumarabic,amssymb, nofootinbib, notitlepage, 11pt, aps, prd]{revtex4-1}
\usepackage{dcolumn}   
\usepackage{bm, bbm, bbding}        
\usepackage{amssymb,amsmath,amsfonts,mathrsfs,graphicx,xcolor,units,yfonts}
\usepackage{txfonts,dsfont,extarrows}
\usepackage[toc,page,title,titletoc,header]{appendix}
\usepackage{setspace, framed, hyperref}

\begin{document}

\title{Friedmann equations from nonequilibrium
thermodynamics of the Universe: A unified formulation for modified gravity}%

\author{David W. Tian}%
\email[]{wtian@mun.ca}
\affiliation{Faculty of Science,  Memorial University, St. John's, Newfoundland, Canada, A1C 5S7}
\author{Ivan Booth}%
\email[]{ibooth@mun.ca}
\affiliation{Department of Mathematics and Statistics, Memorial University, St. John's,  Newfoundland, Canada, A1C 5S7}

\begin{abstract}
Inspired by the Wald-Kodama entropy $S=A/(4G_{\text{eff}})$ where $A$ is the horizon area and $G_{\text{eff}}$ is the effective gravitational coupling strength in modified gravity with field equation $R_{\mu\nu}-Rg_{\mu\nu}/2=$ $8\pi G_{\text{eff}}  T_{\mu\nu}^{\text{(eff)}}$, we
develop a unified and compact formulation in which the Friedmann equations can be derived from thermodynamics of the Universe. The Hawking and Misner-Sharp masses are generalized by replacing Newton's constant $G$ with $G_{\text{eff}}$, and the unified first law of equilibrium thermodynamics is supplemented by a nonequilibrium  energy dissipation term $\mathcal{E}$ which arises from the revised continuity equation of the perfect-fluid effective matter content and is related to the evolution of $G_{\text{eff}}$. By identifying the mass as the total internal energy, the unified first law for the interior and its smooth transit to the apparent horizon yield both Friedmann equations, while the nonequilibrium Clausius relation with entropy production for an isochoric process provides an alternative derivation on the horizon.
We also analyze the equilibrium situation $G_{\text{eff}}=G=\text{constant}$, provide a viability test of the generalized geometric masses, and discuss the continuity/conservation equation. Finally, the general formulation is applied to the FRW cosmology of minimally coupled $f(R)$, generalized Brans-Dicke, scalar-tensor-chameleon, quadratic, $f(R,\mathcal{G})$ generalized Gauss-Bonnet and dynamical Chern-Simons gravity. In these theories we also analyze the $f(R)$-Brans-Dicke equivalence,  find that the chameleon effect causes extra energy dissipation and entropy production, geometrically reconstruct the mass $\rho_m V$ for the physical  matter content, and show the self-inconsistency of $f(R,\mathcal{G})$  gravity in problems involving $G_{\text{eff}}$.\\

\end{abstract}

\maketitle

\section{Introduction}\label{Introduction}

Ever since the discovery of black hole thermodynamics \cite{Black hole mechanics}, physicists have been searching for more and deeper connections between
relativistic gravity and fundamental laws of thermodynamics. One avenue of investigation by Gibbons and Hawking \cite{de Sitte thermodynamics} found that the event
horizon with radius $\ell$ for the de Sitter spacetime also produces Hawking radiation of temperature $1/(2\pi \ell)$.
Jacobson  \cite{Jacobson 1995} further
showed within general relativity (GR) that on any local Rindler horizon, the entropy $S=A/4G$ and the Clausius relation
$TdS=\delta Q$ could reproduce Einstein's field equation, with $\delta Q$ and  $T$ being the
energy flux and the Unruh temperature \cite{Unruh Temperature}.

Besides global and quasilocal black-hole horizons \cite{Hayward trapping horizons, Isolated Horizons Hamiltonian} and the local Rindler horizon, another familiar class of horizons are the various cosmological horizons. Frolov and Kofman \cite{de Sitte inflation} showed that for the flat quasi-de Sitter inflationary universe, $dE=TdS$ yields the Friedmann equation for the
rolling inflaton field, and with metric and entropy perturbations it
reproduces the linearized Einstein equations. By studying the heat flow during an infinitesimal time interval on the apparent horizon of the FRW universe within GR,
Cai and Kim \cite{Cai I} showed that the Clausius thermal relation $TdS=\delta Q=-A\bm\psi$ yields the second Friedmann gravitational equation  with any spatial curvature,
from which the first Friedmann equation can be directly recovered via the continuity/conservation
equation of the perfect-fluid matter content. This work soon attracted much interest,
and cosmology in different dark-energy content and gravity theories came into attention. 

In \cite{Cai II} it was found that extensions of this formulation from GR to $f(R)$ and scalar-tensor theories are quite nontrivial, and the entropy formulas $S=Af_R/4G$ and  $S=Af(\phiup)/4G$ for black-hole horizons prove inconsistent in recovering Friedmann equations.
In the meantime, Eling et al. \cite{Eling Nonequilibrium Thermodynamics} studied nonequilibrium thermodynamics of spacetime and
found that $f(R)$ gravity indeed corresponds to a nonequilibrium description and therefore needs
an entropy production term to balance the energy supply;
the nonequilibrium Clausius relation $\delta Q=T(dS+d_p S)$ with
$S=A f_R/(4G)$ then recovers the Friedmann equations.
This nonequilibrium picture has been widely accepted, and relativistic gravity theories with nontrivial coefficient for $R_{\mu\nu}$ or equivalently $T_{\mu\nu}^{(m)}$ (hence nontrivial gravitational coupling strength $G_{\text{eff}}$) in their field equations always require a nonequilibrium description.
Following \cite{Eling Nonequilibrium Thermodynamics}, Friedmann equations are recovered from nonequilibrium thermodynamics within scalar-tensor gravity with horizon entropy $S=A f(\phiup)/(4G)$ \cite{Cai III-2}. Besides the most typical $f(R)$ \cite{Cai II, Eling Nonequilibrium Thermodynamics} and scalar-tensor \cite{Cai II, Cai III-2} gravity, Friedmann equations from the Clausius relation are also studied in higher-dimensional gravity models like Lovelock gravity \cite{Cai I, Cai III-2} and Gauss-Bonnet gravity \cite{Cai I}.

In the early investigations within modified and alternative theories of gravity,
the standard definition of the Misner-Sharp mass \cite{Misner-Sharp mass}
was used. However, the interesting fact that higher-order geometrical term or
extra physical degrees of freedom beyond GR act like an effective matter content encourages the attempts
to generalize such geometric definitions of mass in modified gravity.
\cite{Misner-Sharp mass III} generalized the Misner-Sharp mass in $f(R)$ gravity,
and also for the FRW universe in the scalar-tensor gravity.
In \cite{mass-like function}, a masslike function was employed
in place of the standard Misner-Sharp mass, so that for $f(R)$  and scalar-tensor gravity the Friedmann equations on the apparent horizon  could be
recovered from the equilibrium Clausius relation $TdS=\delta Q$  without the nonequilibrium correction of \cite{Eling Nonequilibrium Thermodynamics}.
Moreover, the opposite process of \cite{Cai I} to inversely rewrite the Friedmann equations into the thermodynamic relations has been investigated as well. For example, \cite{Cai V} studies such reverse process for GR, Lovelock and Gauss-Bonnet gravity, \cite{Cai IV} for $f(R)$ gravity, \cite{Inverse brane world} for the braneworld scenario, and \cite{scalar f R phi phi2} for generic $f(R\,,\phiup\,,\nabla_\alpha\phiup \nabla^\alpha \phiup)$ gravity.  Also, the field equations of various modified gravity are recast into the form of the Clausius relation in \cite{Clausius modified gravity}. One should carefully distinguish the problem of ``thermodynamics to Friedmann equations'' with ``Friedmann equations to thermodynamics'', to avoid falling into the trap of cyclic logic.

Considering the discreteness of these works following \cite{Cai I} and the not-so-consistent setups of thermodynamic quantities therein, we are pursuing a simpler and more concordant mechanism hiding behind them:
the purpose of this paper is to develop a unified formulation which derives the Friedmann equations from the (non)equilibrium thermodynamics of the FRW universe within all relativistic gravity with field equation $R_{\mu\nu}-Rg_{\mu\nu}/2=8\pi G_{\text{eff}}  T_{\mu\nu}^{\text{(eff)}}$ with a possibly dynamical $G_{\text{eff}}$. These theories include fourth-order modified theories of gravity in the metric approach (as opposed to Palatini) (eg. \cite{Dark Energy II, AA Tian-Booth Paper}) with Lagrangian densities like $\mathscr{L}=f(R)+16\pi G \mathscr{L}_m$ \cite{Example fR}\,, $\mathscr{L}=f(R,\mathcal{G})+16\pi G \mathscr{L}_m$ \cite{Example GaussBonnet second model f(R G)+Lm} ($\mathcal{G}$ denoting the Gauss-Bonnet invariant), $\mathscr{L}=f(R,R_{\mu\nu}R^{\mu\nu}, R_{\mu\alpha\nu\beta}R^{\mu\alpha\nu\beta})+16\pi G \mathscr{L}_m$ \cite{Example Carroll R+ f(R Rc2 Rm2)+2kLm} and quadratic gravity \cite{Example Quardratic gravity second paper}; alternative theories of gravity\footnote{For brevity, we will use the terminology ``modified gravity'' to denote both modified and alternative theories of relativistic gravity without discrimination whenever appropriate.} like Brans-Dicke \cite{Brans Dicke} and scalar-tensor-chameleon \cite{scalar tensor chameleeon} in the Jordan frame;  typical dark-energy models $\mathscr{L}=R+f(\phiup\,,\nabla_\alpha\phiup \nabla^\alpha \phiup)+16\pi G \mathscr{L}_m$ \cite{Dark Energy}, and even
generic mixed models like $\mathscr{L}=f(R\,,\phiup\,,\nabla_\alpha\phiup \nabla^\alpha \phiup)+16\pi G \mathscr{L}_m$ (eg. \cite{scalar f R phi phi2}). All have minimal geometry-matter coupling with isolated matter Lagrangian density $\mathscr{L}_m$. The situation with nonminimal curvature-matter coupling terms \cite{Nonminimal coupling 0, Nonminimal coupling} like $R\mathscr{L}_m$ will not be considered in this paper, although the nonminimal chameleon coupling $\phiup\mathscr{L}_m$ \cite{scalar tensor chameleeon, scalar tensor chameleeon II} in scalar-tensor gravity is still analyzed.

This paper is organized as follows. Sec.~\ref{Preparations and setups} makes necessary preparations by locating the marginally inner trapped horizon as the apparent horizon of the FRW universe, revising the continuity equation for effective perfect fluids, and introducing the energy dissipation term $\mathcal{E}$ for modified gravity with field equation $R_{\mu\nu}-Rg_{\mu\nu}/2=8\pi G_{\text{eff}}  T_{\mu\nu}^{\text{(eff)}}$. In Sec.~\ref{Inside the cosmological apparent horizon}, we generalize the geometric definitions of mass using $G_{\text{eff}}$, supplement the unified first law of thermodynamics into $dE=A\bm\psi+WdV+\mathcal{E}$ by $\mathcal{E}$, and match the transverse gradient of the geometric mass
with the change of total internal energy to directly obtain both Friedmann equations. We continue to study the thermodynamics of the apparent horizon by taking the smooth limit from the interior to the horizon in Sec.~\ref{On the cosmological apparent horizon}, and alternatively obtain the Friedmann equation from the nonequilibrium Clausius relation $T(dS+d_pS)=\delta Q=-(A\bm\psi_t+\mathcal{E})$, where $d_pS$ represents entropy production which is generally nontrivial unless $G_{\text{eff}}=$ constant . After developing the generic theories, Sec.~\ref{Discussion on the unified first law of thermodynamics} provides  a viability test for the generalized geometric masses,
discusses the continuity equation,  and analyzes the  equilibrium case of $G_{\text{eff}}=G=$ constant  with vanishing dissipation $\mathcal{E}=0$  and entropy production $d_pS=0$.
Finally in Sec.~\ref{Examples}, the theory is applied to $f(R)$, generalized Brans-Dicke,  scalar-tensor-chameleon, quadratic, $f(R,\mathcal{G})$ generalized Gauss-Bonnet and dynamical Chern-Simons gravity, with comments on existing treatment in $f(R)$ and scalar-tensor theories.  Throughout this paper, especially for Sec.~\ref{Examples}, we adopt the sign convention  $\Gamma^\alpha_{\delta\beta}=\Gamma^\alpha_{\;\;\,\delta\beta}$,  $R^{\alpha}_{\;\;\beta\gamma\delta}=\partial_\gamma \Gamma^\alpha_{\delta\beta}-\partial_\delta \Gamma^\alpha_{\gamma\beta}\cdots$ and  $R_{\mu\nu}=R^\alpha_{\;\;\mu\alpha\nu}$ with the metric signature $(-,+++)$.


\section{Preparations and setups}\label{Preparations and setups}


\subsection{FRW cosmology and location of the apparent horizon}

The Friedman-Robertson-Walker (FRW) metric provides the most general solution describing a spatially homogeneous and isotropic Universe. It is not just a theoretical construct: it matches with observations. As such it must, a priori,  be a solution of any aspiring modified or alternative theory of gravity \cite{Dark Energy II}. In the comoving coordinates $(t,r,\theta,\varphi)$ the line element reads (eg. \cite{Cai I})
\begin{equation}\label{FRW metric I}
\begin{split}
ds^2\,&= -dt^2+\frac{a(t)^2}{1-kr^2}\,dr^2 + a(t)^2 r^2\,\Big( d\theta^2+\sin^2 \!\theta \,d\varphi^2 \Big)\\
&=\,h_{\alpha\beta}\,dx^\alpha dx^\beta+\Upsilon^2\, \Big( d\theta^2 +\sin^2 \!\theta \,d\varphi^2 \Big)\;,
\end{split}
\end{equation}
where the curvature index  $k$ is normalized to one of $\{-1\,,0\,,+1\}$ which correspond to  closed,
flat and open universes, respectively; the metric function $a(t)$ is the scale factor, which is
an arbitrary function  of the comoving time and is to be determined by the particular gravitational field equations.
$h_{\alpha\beta}\coloneqq\text{diag}[-1\,, \frac{a(t)^2}{1-kr^2}]$  is the transverse two-metric spanned
by  $x^\alpha=(t,r)$, and $\Upsilon\coloneqq a(t)\,r$ is the astrophysical  circumference/areal radius.
Although observations currently support a flat universe with $k=0$,  we will allow for
all three situations $k=\{0\,,\pm1\}$ of spatial homogeneity and isotropy  throughout this paper. 

This solution is spherically symmetric and so in studying its physical and geometric properties it is convenient to work with a null tetrad\footnote{The null tetrad formalism and all Newman-Penrose quantities in use here are adapted to the metric signature $(-,+++)$, which is the preferred convention for quasilocal black hole horizons (see eg.  the Appendix B of \cite{Isolated Horizons Hamiltonian}). Also,
the tetrad can be rescaled by $\ell^\mu\mapsto e^{f}\ell^\mu$ and
$n^\mu\mapsto\,e^{-f}n^\mu$ for an arbitrary function $f$, and consequently
$\theta_{(\ell)}\mapsto e^{f}\theta_{(\ell)}$ and
$\theta_{(n)}\mapsto e^{-f}\theta_{(n)}$.}  adapted to this symmetry:
\begin{equation}\label{Tetrad I}
\begin{split}
\ell^\mu\,=\,\left(\,1\,,\frac{\sqrt{1-kr^2}}{a}\,,0\,,0 \right)
\quad,\quad &n^\mu\,=\,\frac{1}{2}\,\left(\,1\,,-\frac{\sqrt{1-kr^2}}{a}\,,0\,,0 \right)
\quad,\quad m^\mu\,=\,\frac{1}{\sqrt{2}\,\Upsilon}\,\Big(0,0,1,\frac{i}{\sin\!\theta}\Big)\;,
\end{split}
\end{equation}
where the null vectors $\ell^\mu$ and $n^\mu$ have respectively been
adapted to the outgoing and ingoing null directions.
The tetrad obeys the cross normalization $\ell_\mu n^\mu=-1$ and $m_\mu \bar{m}^a=1$, and thus
the inverse metric satisfies $g^{\mu\nu}=-\ell^\mu n^\nu-n^\mu \ell^\nu +m^\mu \bar{m}^\nu + \bar{m}^\mu m^\nu$. In this tetrad,  the outward and inward expansions of radial null flow are found to be
\begin{equation}
\theta_{(\ell)}\,= -\big(\rho_{\text{NP}}+\bar{\rho}_{\text{NP}}\big) \,=\,\frac{2r\dot{a}+2\sqrt{1-kr^2}}{a\,r}
\,=\,2H+2\Upsilon^{-1} \sqrt{1-\frac{k\Upsilon^2}{a^2}}
\end{equation}
and
\begin{equation}
\theta_{(n)}\,=\,\mu_{\text{NP}}+\bar{\mu}_{\text{NP}}\,=\,\frac{r\dot{a}-\sqrt{1-kr^2}}{a\,r}
\,=\,H-\Upsilon^{-1} \sqrt{1-\frac{k\Upsilon^2}{a^2}}\;,
\end{equation}
where $\rho_{\text{NP}} \coloneqq -m^\mu \bar{m}^\nu \nabla_\nu \ell_\mu$ and $\mu_{\text{NP}} \coloneqq \bar{m}^\mu m^\nu\nabla_\nu n_\mu$ are two Newman-Penrose spin coefficients, and
$H$ is Hubble's parameter
\begin{equation}
H\,\coloneqq \,\frac{\dot{a}}{a}\;,
\end{equation}
with the overdot denoting the derivative with respect to the comoving time $t$.
In our universe in which  $\dot{a}>0$ and $H>0$ the outward expansion $\theta_{(\ell)}$ is always positive while $\theta_{(n)}$ can easily be seen to vanish when
\begin{equation}\label{Horizon location}
r_{\text{A}} \,=\,\frac{1}{\sqrt{\dot{a}^2+k}}
\qquad\Leftrightarrow\qquad
\Upsilon_{\text{A}} \,=\,\frac{1}{\sqrt{H^2+\displaystyle\frac{k}{a^2}}}\;.
\end{equation}
On this surface
\begin{equation}
\begin{split}
&\theta_{(\ell)}\,=\,4H>0\;,
\end{split}
\end{equation}
and thus $\Upsilon=\Upsilon_{\text{A}}$ is a marginally inner trapped horizon \cite{Hayward trapping horizons} with $\theta_{(n)}<0$ for $\Upsilon<\Upsilon_{\text{A}}$ and  $\theta_{(n)}>0$ for $\Upsilon>\Upsilon_{\text{A}}$. It is identified as the apparent horizon of the FRW universe\footnote{By the original definition \cite{Hawking Ellis} an apparent horizon is always marginally outer trapped with $\theta_{(\ell)}=0$. However in this paper we follow the more general cosmological vernacular convention which defines an apparent horizon to be either a marginally outer trapped or marginally inner trapped surface. In a contracting universe with $\dot{a}<0$ and $H<0$, however, we would have a more standard marginally outer trapped horizon with
$\theta_{(\ell)}=0$ and $\theta_{(n)}=2H<0 $ at $\Upsilon=\Upsilon_{\text{A}}$.}.
Unlike the cosmological event horizon $\Upsilon_{\text{E}}\,\coloneqq\,a\int_t^\infty a^{-1}dt$ \cite{Event horizon}, which is the horizon of \emph{absolute causality} and relies on the entire future history of the universe, the geometrically defined apparent horizon
$\Upsilon_{\text{A}}$ is the horizon of \emph{relative causality} and is observer-dependent:
if we center our coordinate system on any observer comoving with the universe, then
$r_{\text{A}}$ is the coordinate location of the apparent horizon relative to that observer.
$\Upsilon_{\text{A}}$ is practically more useful and realistic in observational cosmology  as it can be identified by local observations in short duration.
In fact, it has been found that \cite{Apparent not Event horizon 0} for an accelerating universe driven by scalarial  dark energy  with a possibly varying equation of state,  the first and second
laws of thermodynamics hold on $\Upsilon_{\text{A}}$ but break down on $\Upsilon_{\text{E}}$.
Moreover for black holes, Hajicek  \cite{Apparent not Event horizon} has argued that Hawking radiation happens on the apparent horizon rather than the event horizon.
Hence in this paper we will focus on the cosmological apparent horizon $\Upsilon_{\text{A}}$.
Note that  in spherical symmetry  $\Upsilon_{\text{A}}$ can equivalently be specified by setting
$g^{\mu\nu}\partial_\mu \Upsilon\partial_\nu \Upsilon$$=h^{\alpha\beta}\partial_\alpha \Upsilon\partial_\beta \Upsilon=0$,
which locates the hypersurface on which $\partial_\alpha \Upsilon$ becomes a null vector.
Hereafter, quantities related to or evaluated on the apparent horizon
$\Upsilon=\Upsilon_{\text{A}}$ will be highlighted by the subscript \emph{A}.

In some calculations we will find it useful to work with the metric with radial coordinate $\Upsilon$ rather than $r$. To that end note that the total derivative of the physical radius $\Upsilon=a(t)r$ yields
\begin{equation}\label{adr replacement}
adr\,=\,d\Upsilon-H\Upsilon dt\;,
\end{equation}
so the FRW metric Eq.(\ref{FRW metric I}) can be rewritten into
\begin{equation}\label{FRW metric II}
ds^2\,= \left(1-\frac{k\Upsilon^2}{a^2}\right)^{-1}\Bigg(
-\Big(1-\frac{\Upsilon^2}{\Upsilon_{\text{A}}^{2}}\Big) \,dt^2
- 2H\Upsilon \, dtd\Upsilon
+ d\Upsilon^2 \Bigg)+\Upsilon^2 \Big( d\theta^2 +\sin^2 \!\theta \,d\varphi^2 \Big)\;.
\end{equation}
For Eqs.(\ref{FRW metric I}) and (\ref{FRW metric II}), the coordinate singularity $r^2=1/k$ or $\Upsilon^2=a^2/k$  can be removed in the isotropic radial coordinate $\bar{r}$ with $r\coloneqq\bar{r}\,(1+\frac{k\bar{r}^2}{4})^{-1}$.
Following Eq.(\ref{FRW metric II}) and keeping in mind that $t$ is not orthogonal
to $\Upsilon$ in the $(t\,,\Upsilon\,,\theta\,,\varphi)$ coordinates,  the transverse component of the tetrad can be rebuilt as
as
\begin{equation}\label{Tetrad II}
\begin{split}
\ell^\mu\,=\,\left(\, 1\,,H\Upsilon+\sqrt{1-\frac{k\Upsilon^2}{a^2}}\;,0\,,0 \right)
\quad,\quad &n^\mu\,=\,\frac{1}{2}\,\left(\,1\,,H\Upsilon-\sqrt{1-\frac{k\Upsilon^2}{a^2}}\;,0\,,0 \right)\;,
\end{split}
\end{equation}
with which we obtain the same expansion rates $\{\theta_{(\ell)}\,,\theta_{(n)}\}$ and the horizon location $\Upsilon_{\text{A}}$ as from the previous tetrad Eq.(\ref{Tetrad I}).


\subsection{Modified gravity and energy dissipation}\label{Modified gravity and dissipative energy}

For modified theories of relativistic gravity such as $f(R)$, $f(R,\mathcal{G})$ and $f(R,R_{\mu\nu}R^{\mu\nu}, R_{\mu\alpha\nu\beta}R^{\mu\alpha\nu\beta})$ classes of fourth-order gravity, and alternative theories such as Brans-Dicke and generic scalar-tensor-chameleon gravity,
the field equations can be recast into the following compact GR form,
\begin{equation}\label{FieldEqnGRForm}
G_{\mu\nu} \equiv R_{\mu\nu}-\frac{1}{2}Rg_{\mu\nu} \,=\, 8\pi G_{\text{eff}}  \,T_{\mu\nu}^{\text{(eff)}}
 \quad\text{with}\quad T_{\mu\nu}^{\text{(eff)}} \,=\, T_{\mu\nu}^{(m)}+T_{\mu\nu}^{\text{(MG)}}\;,
\end{equation}
where the effective gravitational coupling strength $G_{\text{eff}}$ relies on the specific gravity model and can be directly recognized from the coefficient of the stress-energy-momentum (SEM) density tensor $T_{\mu\nu}^{(m)}$ for the physical matter content, which is defined from extremizing the matter action functional $\delta \mathcal{I}_m=-\frac{1}{2}\int d^4x \sqrt{-g} T_{\mu\nu} \delta g^{\mu\nu}$\,. For example, as will be extensively discussed later in Sec.~\ref{Examples}, we have $G_{\text{eff}}=G/f_R$ for $f(R)$ gravity,  $G_{\text{eff}}=G/\phiup$ for Brans-Dicke,   $G_{\text{eff}}=G/(1+2aR)$ for quadratic gravity, $G_{\text{eff}}=G/(f_R+2Rf_\mathcal{G})$ for $f(R,\mathcal{G})$ generalized Gauss-Bonnet gravity, and $G_{\text{eff}}=G$ for dynamical Chern-Simons gravity.  All terms beyond GR $\big(G_{\mu\nu}=8\pi GT_{\mu\nu}^{(m)}\big)$ have been packed into $G_{\text{eff}}$ and $T_{\mu\nu}^{\text{(MG)}}$, which together with $T_{\mu\nu}^{(m)}$
comprises the total effective SEM tensor $T_{\mu\nu}^{\text{(eff)}}$.
Furthermore, we assume a perfect-fluid-type content, which in the metric-independent form is
\begin{equation}\label{Effective SEM Perfect fluid}
T^{\mu\,\text{(eff)}}_{\;\;\nu}=\text{diag}\,\big[-\rho_{\text{eff}},P_{\text{eff}},P_{\text{eff}},P_{\text{eff}}\big]\quad,\quad
\rho_{\text{eff}}=\rho_m +\rho_{\text{(MG)}} \quad,\quad
P_{\text{eff}}=P_m+P_{\text{(MG)}}\;,
\end{equation}
so that $T^{\mu(m)}_{\;\;\nu}=\text{diag}[-\rho_m,P_m,P_m,P_m]$ and $T^{\mu\,\text{(MG)}}_{\;\;\nu}=\text{diag}[-\rho_{\text{(MG)}},P_{\text{(MG)}},
P_{\text{(MG)}},P_{\text{(MG)}}]$. Here $\rho_m$ and $P_m$ respectively collect the energy densities and pressures of all matter components in the universe, say $\rho_m=\rho_m$$(\text{baryon dust})+\rho_m$$(\text{radiation})+\rho_m(\text{dark energy})+\rho_m(\text{dark matter})+\cdots$ and the same for $P_m$,
while the effects of modified gravity have been encoded into $G_{\text{eff}}$, $\rho_{\text{(MG)}}$ and $P_{\text{(MG)}}$.
For the spatially homogeneous and isotropic FRW universe of maximal spatial symmetry, the coupling strength $G_{\text{eff}}$, the energy densities $\{\rho_{\text{eff}}\,,\rho_m \,,\rho_{\text{(MG)}}\}$ and the pressures $\{P_{\text{eff}}\,,P_m\,,P_{\text{(MG)}}\}$, are all functions of the comoving time $t$ only.

If we take the covariant derivative of the field equation (\ref{FieldEqnGRForm}), then it follows from the contracted Bianchi identities that the generalized stress-energy-momentum conservation $\nabla_\mu G^{\mu}_{\;\;\,\nu}\,=\,0\,=\,8\pi\nabla_\mu \Big(G_{\text{eff}} \,T^{\mu\,\text{(eff)}}_{\;\;\nu} \Big)$ holds for \emph{all} modified gravity.  With respect to the FRW metric Eq.(\ref{FRW metric I}), only the $t$-component of this conservation equation is nontrivial and leads to the universal relation
\begin{equation}\label{Generalized continuity eqn}
 \dot{\rho}_{\text{eff}}
+ 3 H\,\Big(\rho_{\text{eff}}+P_{\text{eff}}  \Big)   \,=\,-
\frac{\dot{G}_{\text{eff}}}{G_{\text{eff}}} \, \rho_{\text{eff}}\;,
\end{equation}
which serves as the generalized continuity equation for the perfect fluid of Eq.(\ref{Effective SEM Perfect fluid}). Compared with the continuity equation of a cosmological perfect fluid $\dot{\rho}_m
+ 3 H\big(\rho_m+P_m  \big)=0$ within GR, the extra term $-(\dot{G}_{\text{eff}}/G_{\text{eff}}) \, \rho_{\text{eff}}$ shows up in Eq.(\ref{Generalized continuity eqn}) to balance the energy flow.  Since it has the same dimension as the effective density flow $\dot{\rho}_{\text{eff}}$, we introduce the following differential energy  by multiplying $Vdt=\frac{4}{3}\pi\, \Upsilon^3\,dt$ to it,
\begin{equation}\label{Dissipation energy density}
\mathcal {E}\,\coloneqq \,-\frac{4}{3}\pi\, \Upsilon^3\frac{\dot{G}_{\text{eff}}}{G_{\text{eff}}} \, \rho_{\text{eff}}\; dt\;.
\end{equation}
and call it the term of \emph{nonequilibrium energy dissipation}. Note that at this stage in Eq.(\ref{Dissipation energy density}) for $\mathcal{E}$, the $\frac{4}{3}\pi\, \Upsilon^3\, \rho_{\text{eff}}$ should not be combined into some kind of physically defined mass $V\rho_{\text{eff}}=\mathcal{M}_{\text{eff}}$
as its meaning is not clear yet (this is just an issue for security to avoid cyclic logic).

$\mathcal {E}$ is related to the temporal evolution of $G_{\text{eff}}$ and its coupling to $\rho_{\text{eff}}$.  Whether $\mathcal {E}$ drives the evolution of $G_{\text{eff}}$ or contrarily is produced by the evolution of $G_{\text{eff}}$ is however not yet certain. Also, as will be seen later, $\mathcal {E}$ plays an important role below in supplementing the unified first law of equilibrium thermodynamics and calculating the entropy production.
 .


\section{Thermodynamics inside the apparent horizon}\label{Inside the cosmological apparent horizon}

For the FRW universe as a solution to the generic field equation (\ref{FieldEqnGRForm}),
we substitute the effective  gravitational coupling strength $G_{\text{eff}}$ for Newton's constant $G$ and thus generalize the Hawking mass $M_{\text{Hk}}$  \cite{Hawking mass} for  twist-free spacetimes into
\begin{equation}\label{Hawking mass}
\begin{split}
M_{\text{Hk}}\,\coloneqq\,&\frac{1}{4\pi G_{\text{eff}}} \left( \int  \frac{dA}{4\pi} \right)^{\frac{1}{2}}\int \Big( -\Psi_2-\sigma_{\text{NP}} \lambda_{\text{NP}} +\Phi_{11}+\Lambda_{\text{NP}} \Big)\,dA\\
\equiv\,&\frac{1}{4\pi G_{\text{eff}}} \left( \int  \frac{dA}{4\pi} \right)^{\frac{1}{2}} \Bigg( 2\pi-\int  \rho_{\text{NP}}\,\mu_{\text{NP}}\,dA  \Bigg)\;.
\end{split}
\end{equation}
Since we are dealing with spherical symmetry, $M_{\text{Hk}}$ can  equivalently be written as
\begin{equation}\label{Misner-Sharp mass}
 M_{\text{MS}}\,\coloneqq \,\frac{\Upsilon }{2G_{\text{eff}}} \,
 \Big( 1- h^{\alpha\beta}\partial_\alpha \Upsilon\partial_\beta \Upsilon\Big)\;,
\end{equation}
which similarly generalizes the Misner-Sharp mass $M_{\text{MS}}$ \cite{Misner-Sharp mass}. As will be shown later in Sec.~\ref{A viability test of the extended Hawking and Misner Sharp masses}, the geometric definitions Eqs.(\ref{Hawking mass}) and (\ref{Misner-Sharp mass})
fully reflect the spirit of geometrodynamics that the effective matter content $\rho_{\text{eff}}=\rho_m +\rho_{\text{(MG)}}$ curves the space homogeneously and isotropically through the field equation (\ref{FieldEqnGRForm}) to form the FRW universe. Moreover, the Misner-Sharp mass of black holes in Brans-Dicke gravity with $G_{\text{eff}}=1/\phiup$ has been found to satisfy Eq.(\ref{Misner-Sharp mass}) \cite{BH Mass Brans Dicke}, which also encourages us to make the extensions in Eqs.(\ref{Hawking mass}) and (\ref{Misner-Sharp mass}). Note that the Hawking and Misner-Sharp masses restrict their attentions to the mass of the matter content and do not include the energy of gravitational field.

With $\Psi_2=\sigma_{\text{NP}} =\lambda_{\text{NP}}=0$, $\Phi_{11}=-\Big(\dot{H}- \frac{k}{a^2}\Big)/4$, $\Lambda_{\text{NP}}=\Big(\dot H+2H^2+\frac{k}{a^2}\Big)/4$ or $\rho_{\text{NP}}\,\mu_{\text{NP}}=-\theta_{(\ell)}\theta_{(n)}/4$ in the tetrad Eq.(\ref{Tetrad I}), and $h^{\alpha\beta}$$=\text{diag}[-1\,, \frac{a^2}{1-kr^2}]$ for the transverse two-metric in Eq.(\ref{FRW metric I}), either Eq.(\ref{Hawking mass}) and Eq.(\ref{Misner-Sharp mass}) yield that the mass enveloped by a standard sphere of physical radius $\Upsilon$ in the FRW universe is
\begin{equation}\label{mass}
M\,=\,\frac{\Upsilon^3}{2G_{\text{eff}}}\left( H^2+\frac{k}{a^2}\right)\;.
\end{equation}
Immediately, the total derivative or the transverse gradient of $M=M(t,r)$ is
\begin{eqnarray}
dM\,
&=&\frac{ \Upsilon^3 H}{2G_{\text{eff}}}\,\left(2\dot{H}+3H^2+\frac{k}{a^2}   \right)\,dt
+\frac{3\Upsilon^2}{2G_{\text{eff}}}\,\left(H^2+\frac{k}{a^2}   \right)\,adr-
\frac{\Upsilon^3 \dot{G}_{\text{eff}}}{2G_{\text{eff}}^2}\left( H^2+\frac{k}{a^2}\right)\,dt \label{dM inside t r}\\
&=&\frac{ \Upsilon^3 H}{G_{\text{eff}}}\,\left(\dot{H}-\frac{k}{a^2}   \right)\,dt
+\frac{3\Upsilon^2}{2G_{\text{eff}}}\,\left(H^2+\frac{k}{a^2}   \right)\,d\Upsilon-
\frac{\Upsilon^3 \dot{G}_{\text{eff}}}{2G_{\text{eff}}^2}\left( H^2+\frac{k}{a^2}\right)\,dt \label{dM inside t Upsilon}\;,
\end{eqnarray}
where Eq.(\ref{adr replacement}) has been used to reexpress Eq.(\ref{dM inside t r}) into Eq.(\ref{dM inside t Upsilon}) in terms of the $(t\,,\Upsilon)$ normal coordinates.

Hayward derived a unified first law of  equilibrium thermodynamics \cite{Hayward Unified first law, Hayward Dynamic black-hole entropy} for the differential element of energy change within GR under spherical symmetry, which however will be taken as a \emph{first principle} in our work. For modified gravity of the form Eq.(\ref{FieldEqnGRForm}),
 we  supplement Hayward's result by
the energy dissipation term $\mathcal{E}$ introduced in Eq.(\ref{Dissipation energy density}), so that the change of energy along the outgoing null normal $\ell^\mu$ across a sphere of radius $\Upsilon$ with surface area $A=4\pi \Upsilon^2$ and volume $V=4\pi \Upsilon^3/3$ is
\begin{eqnarray}\label{Unified first law}
dE\,=\, A\bm\psi+W dV+\mathcal {E}\;,
\end{eqnarray}
where  the covector invariant $\bm\psi$ is the energy/heat flux density, the scalar  invariant $W$ is the work density, and $WdV=WAd\Upsilon$. We formally inherit the original definitions of $\{\bm\psi\,,W\}$ \cite{Hayward Unified first law} but make use of the total effective SEM tensor $T_{\mu\nu}^{\text{(eff)}}$ rather than just $T_{\mu\nu}^{(m)}$ as in GR:
\begin{equation}
\psi_\alpha\,\coloneqq\,T_{\alpha\,\text{(eff)}}^{\;\;\,\beta}\,\partial_\beta \Upsilon+W\,\partial_\alpha \Upsilon
\quad\text{with}\quad W\,\coloneqq -\frac{1}{2}\,T^{\alpha\beta}_{\text{(eff)}}\,h_{\alpha\beta}\;,
\end{equation}
where $T_{\alpha\beta}^{\text{(eff)}}$ denote the components of  $T_{\mu\nu}^{\text{(eff)}}$ along the transverse directions. Note that the definitions of $\bm\psi$ and $W$ also guarantee that they are independent of the coordinate systems or observers and the choice of metric signature. Moreover, with the matter content of effective perfect fluid assumed in Eq.(\ref{Effective SEM Perfect fluid}), $\bm\psi$ and $W$ explicitly become
\begin{equation}
W=\frac{1}{2}\,\Big(\rho_{\text{eff}}-P_{\text{eff}}\Big) \quad\text{and}\quad
\end{equation}
\begin{equation}
\begin{split}
\bm\psi\;&=\;-\frac{1}{2} \,\Big(\rho_{\text{eff}}+P_{\text{eff}}\Big)\,H\Upsilon\,dt
+ \frac{1}{2} \,\Big(\rho_{\text{eff}}+P_{\text{eff}}\Big)\,a\,dr\\
&=\;- \;\;\, \Big(\rho_{\text{eff}}+P_{\text{eff}}\Big)\,H\Upsilon\,dt
+ \frac{1}{2} \,\Big(\rho_{\text{eff}}+P_{\text{eff}}\Big)\,d\Upsilon\;,
\end{split}
\end{equation}
where $W$ no longer preserves the generalized energy conditions\footnote{For the field equation (\ref{FieldEqnGRForm}) along with
$R =-8\pi G_{\text{eff}}\, T^{\text{(eff)}}$ and $R_{\mu\nu} = 8\pi G_{\text{eff}}\, \big(T_{\mu\nu}^{\text{(eff)}}-\frac{1}{2} g_{\mu\nu}T^{\text{(eff)}} \big)$\,,
the Raychaudhuri equations (\cite{Hawking Ellis} or the appendix of \cite{AA Tian-Booth Paper}) imply the following null, weak  and strong energy conditions (abbreviated into NEC, WEC and SEC respectively):
\begin{equation*}
G_{\text{eff}}T_{\mu\nu}^{\text{(eff)}}\,\ell^\mu \ell^\nu \,\geq\,0\quad (\text{NEC})\quad,\quad
G_{\text{eff}}T_{\mu\nu}^{\text{(eff)}}\,u^\mu u^\nu \,\geq\,0\quad (\text{WEC})\quad,\quad
G_{\text{eff}}T_{\mu\nu}^{\text{(eff)}}u^\mu u^\nu\,\geq\,\frac{1}{2}\, G_{\text{eff}}T^{\text{(eff)}}u_\mu u^\mu
\quad (\text{SEC})\;,
\end{equation*}
where $u_\mu u^\mu=-1$  in the SEC for the metric signature $(-,+++)$ used in this paper. All energy conditions require $G_{\text{eff}}\,\big(\rho_{\text{eff}}-P_{\text{eff}}\big) \geq 0$ for the effective matter content Eq.(\ref{Effective SEM Perfect fluid}).} as opposed to the
situation of GR \cite{Hayward Unified first law} unless $G_{\text{eff}}$ is positive definite.
Hence,
the unified first law Eq.(\ref{Unified first law}) leads to
\begin{eqnarray}
dE\,&=&\,-A \Upsilon  H \,P_{\text{eff}}\,dt+A\,\rho_{\text{eff}}\, adr
-\frac{4}{3}\pi\, \Upsilon^3\frac{\dot{G}_{\text{eff}}}{G_{\text{eff}}} \, \rho_{\text{eff}} \, dt \label{dE inside t r}\\
&=&-A\,\Big(\rho_{\text{eff}}+P_{\text{eff}}\Big)\,H\Upsilon\,dt
+ A\,\rho_{\text{eff}}\,d\Upsilon-\frac{4}{3}\pi\, \Upsilon^3\frac{\dot{G}_{\text{eff}}}{G_{\text{eff}}} \, \rho_{\text{eff}} \,dt \label{dE inside t Upsilon}\;.
\end{eqnarray}

Hence, by identifying the geometrically defined mass $M$ as the \emph{total internal energy},
matching the coefficients of $dt$ and $dr$ in Eqs.(\ref{dM inside t r}) and (\ref{dE inside t r})
or the coefficients of $dt$ and $d\Upsilon$ in Eqs.(\ref{dM inside t Upsilon}) and (\ref{dE inside t Upsilon}),
we obtain
\begin{equation}\label{Friedmann eqn 1st}
H^2+\frac{k}{a^2}\,=\, \frac{8\pi G_{\text{eff}}}{3}\,\rho_{\text{eff}} \quad\text{and}\quad
\end{equation}
\begin{equation}\label{Friedmann eqns 2nd}
\dot H-\frac{k}{a^2}=-4\pi G_{\text{eff}}\,\Big(\rho_{\text{eff}}+P_{\text{eff}}\Big)\qquad\text{or}\qquad
2\dot H+3H^2+\frac{k}{a^2}\,= -8\pi G_{\text{eff}} P_{\text{eff}}\;,
\end{equation}
where we have recognized the last term in  Eqs.(\ref{dM inside t r}) and (\ref{dM inside t Upsilon}) for $dM$ equal to the  dissipation  $\mathcal{E}$ in $dE$ as they are both relevant to the evolution of $G_{\text{eff}}$.

In fact, by substituting the FRW metric Eq.(\ref{FRW metric I}) into the field equation (\ref{FieldEqnGRForm}), it can be verified that Eqs.(\ref{Friedmann eqn 1st}) and (\ref{Friedmann eqns 2nd}) are exactly the first and the second Friedmann equations governing the dynamics of the scale factor $a(t)$ for the FRW cosmology.
Hence, the gravitational equations (\ref{Friedmann eqn 1st}) and (\ref{Friedmann eqns 2nd}) have been derived from the unified first law of  nonequilibrium thermodynamics $dE=A\bm\psi+WdV +\mathcal {E}$ instead of the field equation (\ref{FieldEqnGRForm}), and this is not a result of cyclic logic as Eqs.(\ref{Friedmann eqn 1st}) and (\ref{Friedmann eqns 2nd}) are preassumed as unknown. By the way, for the two versions of the second Friedmann equation in Eq.(\ref{Friedmann eqns 2nd}), the former is generally more preferred than the latter, because the former directly reflects the evolution of the Hubble parameter $H$ (especially for $k=0$ of the observed universe), and in numerical simulations the values of $\dot H$ and $H^2$ can differ dramatically (eg. \cite{de Sitte inflation} with $H^2\gg \dot H$) and thus be problematic to work with when put together.

Once one of the Friedmann equations is known, the other one can be obtained using the continuity equation (\ref{Generalized continuity eqn}). For example, taking the time derivative of the first Friedmann equation $H^2+k/a^2=8\pi G_{\text{eff}}\,\rho_{\text{eff}}/3$,
\begin{equation}
2H\,\Big( \dot H-\frac{k}{a^2}\Big)\,=\,\frac{8\pi }{3}\,\Big( \dot{G}_{\text{eff}} \rho_{\text{eff}}+ G_{\text{eff}} \,\dot{\rho}_{\text{eff}} \Big)\;,
\end{equation}
and applying the continuity equation
\begin{equation*}
\dot{G}_{\text{eff}}\, \rho_{\text{eff}}+ G_{\text{eff}} \,\dot{\rho}_{\text{eff}}+
3G_{\text{eff}} \,H\,\Big(\rho_{\text{eff}}+P_{\text{eff}}  \Big)  \,=\,0\;,
\end{equation*}
one recovers the second Friedmann equation
$\displaystyle \dot H-k/a^2=-4\pi G_{\text{eff}}\,(\rho_{\text{eff}}+P_{\text{eff}})$.
Inversely, integration of the second Friedmann equation with the continuity equation leads to the first Friedmann equation by
neglecting an integration constant or otherwise treat it as a cosmological constant \cite{Cai I} and incorporate it into $\rho_{\text{eff}}$.


\section{Thermodynamics On the apparent horizon}\label{On the cosmological apparent horizon}

Having derived the Friedmann equations from the thermodynamics of the FRW universe inside the apparent horizon
$\Upsilon< \Upsilon_{\text{A}}$, we will continue to study this thermodynamics-gravity correspondence on the horizon $\Upsilon= \Upsilon_{\text{A}}$, and in the meantime require consistency between the interior and the horizon. In fact,
existing papers about this problem almost exclusively focus on the horizon alone \cite{Cai I, Cai II, Cai III-2, mass-like function}, as a companion to the thermodynamics of black-hole and Rindler horizons.
In this section,  the apparent horizon $\Upsilon= \Upsilon_{\text{A}}$ will be studied
via two methods: (1) Following Sec.\ref{Inside the cosmological apparent horizon}, applying the nonequilibrium unified first law $dE=A\bm\psi +WdV+\mathcal {E}$ and $dE=dM$ in the smooth limit $\Upsilon \to \Upsilon_{\text{A}}$; (2) Using the nonequilibrium Clausius relation $T(dS+d_pS)= \delta Q=-(A\bm\psi+\mathcal{E})$ with entropy production $d_pS$ and the continuity equation (\ref{Generalized continuity eqn}).

\subsection{Method 1: Unified first law and $dE\hat{=}dM$}\label{Method 1 Unified first law}

As shown by Eq.(\ref{Horizon location}) in Sec.\ref{Preparations and setups}, the cosmological apparent horizon, in this case a marginally inner trapped horizon of the expanding FRW universe
locates at
$\Upsilon_{\text{A}}=1/\sqrt{H^2+k/a^2}$, and according to Eq.(\ref{mass}),
the mass within the horizon is
$M_{\text{A}}=  \Upsilon_{\text{A}}/(2G_{\text{eff}})$.
Following Sec.~\ref{Modified gravity and dissipative energy} and taking the smooth limit
$\Upsilon\to \Upsilon_{\text{A}}$ from the interior to the horizon,  Eqs.(\ref{dM inside t r}) and (\ref{dE inside t r}) yield in the $(t\,,r)$ comoving transverse coordinates that
\begin{eqnarray}
d M\,&\hat{=}&\,\frac{ \Upsilon^3_{\text{A}} H}{2G_{\text{eff}}}\,\left(2\dot{H}+3H^2
+\frac{k}{a^2}   \right)\,dt
+\frac{3a}{2G_{\text{eff}}}\,dr
-\frac{\Upsilon_{\text{A}} \dot{G}_{\text{eff}}}{2G_{\text{eff}}^2}\,dt \label{dE=dM horizon I}\\
dE\,&\hat{=}&-A_{\text{A}} \Upsilon_{\text{A}}  H \,P_{\text{eff}}\,dt
+A_{\text{A}}\,\rho_{\text{eff}}\, adr
-\frac{4}{3}\pi\, \Upsilon_{\text{A}}^3\frac{\dot{G}_{\text{eff}}}{G_{\text{eff}}} \, \rho_{\text{eff}}\, dt\label{dE=dM horizon II}\;,
\end{eqnarray}
while Eqs.(\ref{dM inside t Upsilon}) and (\ref{dE inside t Upsilon})
in the $(t\,,\Upsilon)$ coordinates  give rise to
\begin{eqnarray}
d M\;\hat{=}\;\frac{ \Upsilon_{\text{A}}^3 H}{G_{\text{eff}}}\,\left(\dot{H}-\frac{k}{a^2}   \right)\,dt
+\frac{3}{2G_{\text{eff}}}\,d\Upsilon
&-&\frac{\Upsilon_{\text{A}} \dot{G}_{\text{eff}}}{2G_{\text{eff}}^2} \,dt \label{dE=dM horizon III}\\
dE\,\hat{=}-A_{\text{A}}\,\Big(\rho_{\text{eff}}+P_{\text{eff}}\Big)\,H\Upsilon_{\text{A}}\,dt
+ A_{\text{A}}\,\rho_{\text{eff}}\,d\Upsilon
&-&\frac{4}{3}\pi\, \Upsilon_{\text{A}}^3\frac{\dot{G}_{\text{eff}}}{G_{\text{eff}}} \, \rho_{\text{eff}} \,dt\label{dE on horizon}\;,
\end{eqnarray}
where the symbol $\hat{=}$ will be employed hereafter to denote ``equality on the apparent horizon'', a standard denotation widely used for equality on quasilocal black-hole horizons (eg. \cite{Isolated Horizons Hamiltonian}). Note that for the $dr$ components in Eqs.(\ref{dE=dM horizon I}) and (\ref{dE=dM horizon II}) as well as the $d\Upsilon$ components in Eqs.(\ref{dE=dM horizon III}) and (\ref{dE on horizon}), one just needs to evaluate their coefficients in the limit  $\Upsilon \to \Upsilon_{\text{A}}$; although both horizon radii  $r_{\text{A}}={r}_{\text{A}}(t)$ and $\Upsilon_{\text{A}}=\Upsilon_{\text{A}}(t)$ are functions of $t$ according to Eq.(\ref{Horizon location}), the differentials $dr$ and $d\Upsilon$ should not be replaced by $\dot{r}_{\text{A}}dt$ and $\dot{\Upsilon}_{\text{A}}dt$ for   $\Upsilon \to \Upsilon_{\text{A}}$,  because the horizon is not treated as a thermodynamical system alone by itself.
As expected, in the limit $\Upsilon \to \Upsilon_{\text{A}}$ the equality $dM\,\hat{=}\,dE$ recovers the Friedmann equations again,
\begin{equation*}
H^2+\frac{k}{a^2}\;\hat{=}\; \frac{8\pi G_{\text{eff}}}{3}\,\rho_{\text{eff}}\quad\text{and}\quad
\dot H-\frac{k}{a^2}\,\hat{=}-4\pi G_{\text{eff}}\,\Big(\rho_{\text{eff}}+P_{\text{eff}}\Big)
\quad\text{or}\quad2\dot H+3H^2+\frac{k}{a^2}\,\hat{=} -8\pi G_{\text{eff}} P_{\text{eff}}\;.
\end{equation*}
Specifically note note from Eqs.(\ref{dE=dM horizon III}) and (\ref{dE on horizon})  that on the horizon the dissipation term satisfies
\begin{equation}\label{horizon dissipation replacement}
\frac{4}{3}\pi\, \Upsilon_{\text{A}}^3\frac{\dot{G}_{\text{eff}}}{G_{\text{eff}}} \, \rho_{\text{eff}}\;\hat{=}\;
\frac{1}{2}\Upsilon_{\text{A}}\frac{\dot{G}_{\text{eff}}}{G_{\text{eff}}^2}\;,
\end{equation}
which, without being further simplified, will be used in the next subsection to reduce the expression of the on-horizon  entropy production.


\subsection{Method 2: Nonequilibrium Clausius relation}\label{Method 2 Clausius relation}

The modified theories of gravity under our consideration with the field equation  (\ref{FieldEqnGRForm}) are all diffeomorphism invariant,
and therefore we can obtain the Wald-Kodama  dynamical  entropy of the FRW apparent horizon  by Wald's Noether-charge method \cite{Wald entropy, Wald entropy II, Hayward Dynamic black-hole entropy} as
\begin{equation}\label{entropy Wald-Kodama}
S\,\coloneqq \, \int   \frac{dA}{4G_{\text{eff}}}\;\hat{=}\;\frac{A_{\text{A}}}{4G_{\text{eff}}} \;\hat{=}\;\frac{\pi \Upsilon_{\text{A}}^2 }{G_{\text{eff}}}\;,
\end{equation}
with $G_{\text{eff}}=G_{\text{eff}}(t)$. In fact,
the field equations of modified and alternative gravity have been deliberately rearranged
into the form of Eq.(\ref{FieldEqnGRForm}) with an effective gravitational coupling strength $G_{\text{eff}}$ to facilitate the definition of the horizon entropy Eq.(\ref{entropy Wald-Kodama}).
Moreover, the absolute temperature of the horizon is assumed  to be \cite{Cai I}
\begin{equation}\label{T}
T\,\equiv \,\frac{1}{2 \pi \Upsilon_{\text{A}}}\;,
\end{equation}
which agrees with the temperature of the semiclassical thermal spectrum \cite{Temperature tuneling} for the matter tunneling into the region $\Upsilon<\Upsilon_{\text{A}}$ from the exterior $\Upsilon>\Upsilon_{\text{A}}$, as measured by a Kodama observer using the line element Eq.(\ref{FRW metric II}). In fact, if the dynamical surface gravity \cite{Dynamical surface gravity} for the FRW spacetime is defined as $ \kappa\coloneqq -\frac{1}{2}\partial_\Upsilon \Xi$ with $\Xi\coloneqq h^{\alpha\beta}\partial_\alpha \Upsilon\partial_\beta \Upsilon\equiv 1-\Upsilon^2\big(H^2+\frac{k}{a^2}\big)=1-\Upsilon^2/\Upsilon_{\text{A}}^2$, then $\displaystyle \kappa=\Upsilon/\Upsilon_{\text{A}}^2\,\hat{=}\,1/\Upsilon_{\text{A}}$ and the temperature ansatz Eq.(\ref{T}) satisfies $T=\kappa/(2\pi)$. This formally matches the Hawking temperature of (quasi-)stationary black holes in terms of the traditional definition of surface gravity
\cite{Black hole mechanics} based on Killing vectors and Killing horizons.
Hence it follows from Eqs.(\ref{entropy Wald-Kodama}) and (\ref{T}) that
\begin{equation}\label{TdS}
{T}d{S}\;\hat{=}\;  \frac{ \dot{\Upsilon}_{\text{A}}}{G_{\text{eff}}}dt
-\frac{1}{2}\Upsilon_{\text{A}}\frac{\dot{G}_{\text{eff}}}{G_{\text{eff}}^2} dt
\qquad\text{with}\qquad
\dot{\Upsilon}_{\text{A}}\,=\,-H \Upsilon_{\text{A}}^3 \,\left( \dot{H}-\frac{k}{a^2}\right)\;.
\end{equation}

Assuming that at the moment $t=t_0$ the apparent horizon locates at $\Upsilon_{\text{A}0}$, then during the infinitesimal time interval $dt$ the horizon will move to\footnote{The second Friedmann equation (\ref{Friedmann eqns 2nd}) can be rewritten into the evolution equation for  the apparent-horizon radius $\Upsilon_{\text{A}}$:
\begin{equation*}
\dot{\Upsilon}_{\text{A}}\,=\,4\pi\, H \Upsilon_{\text{A}}^3  G_{\text{eff}}\,\big(\rho_{\text{eff}}+P_{\text{eff}}\big) \,,
\end{equation*}
which shows that for an expanding universe ($H>0$), $\Upsilon_{\text{A}}$ can be either expanding, contracting or even static, depending on the values of $G_{\text{eff}}$ and the effective equation of state parameter $w_{\text{eff}}=P_{\text{eff}}/\rho_{\text{eff}}$.} $\Upsilon_{\text{A}0} +\dot{\Upsilon}_{\text{A}0} dt$. In the meantime, for the \emph{isochoric} process  ($d\Upsilon=0$) for the  volume of constant radius $\Upsilon_{\text{A}0} $, the amount of energy across the horizon $\Upsilon=\Upsilon_{\text{A}0} $ during this $dt$ is just $dE\,\hat{=}\,A_{\text{A}}\bm\psi_t+\mathcal{E}_{\text{A}}$ evaluated at $t=t_0$, as has been calculated in  Eq.(\ref{dE on horizon}) with the $d\Upsilon$ component removed.


Compare $dE\,\hat{=}\,A_{\text{A}}\bm\psi_t+\mathcal{E}_{\text{A}}$ with Eq.(\ref{TdS}),
and it turns out the Clausius relation $TdS\,\hat{=}\,\delta Q\,\hat{=}-dE$
for equilibrium thermodynamics does not hold.  To balance the energy change, we have to introduce
an extra  entropy production term $d_pS$ \cite{Eling Nonequilibrium Thermodynamics} (subscript $p$ being short for ``production'') so that
\begin{equation}\label{TdS nonequi}
\begin{split}
TdS+Td_p S\; \hat{=}-dE\;\;\hat{=}-  \big(A_{\text{A}}\bm\psi_t+\mathcal{E}_{\text{A}}\big)\;.
\end{split}
\end{equation}
Hence, it follows from Eqs.(\ref{dE on horizon}) and (\ref{TdS})  that
\begin{equation}\label{TdS TdpS}
\begin{split}
Td_p S\; &\hat{=}\;-TdS- A_{\text{A}}\bm\psi_t-\mathcal{E}_{\text{A}}\\
&\hat{=}\;-\left(\frac{ \dot{\Upsilon}_{\text{A}}}{G_{\text{eff}}}dt+  A_{\text{A}}\bm\psi\right)
+\frac{1}{2}\Upsilon_{\text{A}}\frac{\dot{G}_{\text{eff}}}{G_{\text{eff}}^2} dt-\mathcal{E}_{\text{A}}\\
&\hat{=}\;-\left(\frac{ \dot{\Upsilon}_{\text{A}}}{G_{\text{eff}}} -A_{\text{A}}\,\big(\rho_{\text{eff}}+P_{\text{eff}}\big)\,H\Upsilon_{\text{A}}\right)\,dt
+\frac{1}{2}\Upsilon_{\text{A}}\frac{\dot{G}_{\text{eff}}}{G_{\text{eff}}^2}
+\frac{4}{3}\pi\, \Upsilon_{\text{A}}^3\frac{\dot{G}_{\text{eff}}}{G_{\text{eff}}} \, \rho_{\text{eff}}\, dt\;.
\end{split}
\end{equation}
We have combined the $\dot{\Upsilon}_{\text{A}}$ component of $TdS$ in Eq.(\ref{TdS}) with $A_{\text{A}}\bm\psi_t$\,, which reproduces the second Friedmann equation
\begin{equation}\label{TdS Friedmann}
\frac{ \dot{\Upsilon}_{\text{A}}}{G_{\text{eff}}} - A_{\text{A}}\,\big(\rho_{\text{eff}}+P_{\text{eff}}\big)\,H\Upsilon_{\text{A}} \;\hat{=}\;0
\qquad\Rightarrow\qquad
\dot H-\frac{k}{a^2}\,\hat{=}-4\pi G \Big(\rho_{\text{eff}}+P_{\text{eff}}\Big)\;,
\end{equation}
while the $\dot{G}_{\text{eff}}$ component of $TdS$ in Eq.(\ref{TdS}) and the energy dissipation $\mathcal{E}_{\text{A}}$ add up together and
give rise to the entropy production
\begin{equation}\label{entropy production}
T d_p S\;\hat{=}\;\frac{1}{2}\,\Upsilon_{\text{A}}\,\frac{\dot{G}_{\text{eff}}}{G_{\text{eff}}^2}\, dt+\frac{4}{3}\pi\, \Upsilon_{\text{A}}^3\frac{\dot{G}_{\text{eff}}}{G_{\text{eff}}} \, \rho_{\text{eff}}\, dt\qquad\text{and}\qquad
d_p {S}\;\hat{=}\;
\pi \Upsilon_{\text{A}}^2\frac{\dot{G}_{\text{eff}}}{G_{\text{eff}}^2} \, dt
+\frac{8}{3}\pi^2\, \Upsilon_{\text{A}}^4\frac{\dot{G}_{\text{eff}}}{G_{\text{eff}}} \, \rho_{\text{eff}}\, dt\;.
\end{equation}

Hence, for the Wald-Kodama dynamical entropy Eq.(\ref{entropy Wald-Kodama}), $TdS$ manifests its effects in two aspects: the  $\dot{\Upsilon}_{\text{A}}$ bulk term is the equilibrium part related to the expansion of the universe and the apparent horizon, while the $\dot{G}_{\text{eff}}$ term is the nonequilibrium part associated to the evolution of the coupling strength. The former balances the energy flux $A\bm\psi_t$ and leads to the Friedmann equation (\ref{TdS Friedmann}), while the latter, together with the generic energy dissipation $\mathcal{E}$ evaluated on the horizon, constitute the two sources shown up in Eq.(\ref{entropy production}) responsible for the entropy production.

As discussed before in Sec.~\ref{Inside the cosmological apparent horizon}, the first Friedmann equation $H^2+k/a^2\,\hat{=}\,8\pi G_{\text{eff}}\,\rho_{\text{eff}}/3$
can be obtained from Eq.(\ref{TdS Friedmann}) with the help of the continuity equation (\ref{Generalized continuity eqn}). For the consistency between the horizon and the interior in the relation $dE\,\hat{=}\,dM\,\hat{=}-T(dS+d_p S)$,
we have adjusted the thermodynamic sign convention into $T(dS+d_p S)\,\hat{=}\,\delta Q\,\hat{=}-dE\,\hat{=}-(A_{\text{A}}\bm\psi_t+\mathcal{E}_{\text{A}})$.

In this paper, following the spirit of \cite{Eling Nonequilibrium Thermodynamics}, primarily we call the modified gravity an equilibrium or nonequilibrium theory from the thermodynamic point of view depending on whether the equilibrium Clausius relation $TdS\,\hat{=}\,\delta Q\,\hat{=}-dE\,\hat{=}-A_{\text{A}}\bm\psi_t$ or its nonequilibrium extension with entropy production $TdS+Td_p S \,\hat{=}-dE\,\hat{=}-  \big(A_{\text{A}}\bm\psi_t+\mathcal{E}_{\text{A}}\big)$
works on the apparent horizon. Moreover,
Eq.(\ref{entropy production}) clearly shows that both sources for the nonequilibrium entropy-production $d_pS$  trace back  to the dynamics/evolution of $G_{\text{eff}}$. Hence, we further regard all those quantities containing $\dot{G}_{\text{eff}}$ as nonequilibrium, such as the energy dissipation element introduced in Eq.(\ref{Dissipation energy density}).  In the same sense, $TdS$ itself in Eq.(\ref{TdS}) is no longer a thermodynamical  quasistationary expression, and we regard its $\dot{\Upsilon}_{\text{A}}$ bulk  component   as equilibrium, while its $\dot{\Upsilon}_{\text{A}}$ component as  nonequilibrium. This way, the thermodynamic terminology ``nonequilibrium'' and ``equilibrium'' in our usage throughout this paper have been clarified.

Eq.(\ref{entropy production}) demonstrates that the entropy production effect  is  generally unavoidable in modified  gravity unless $G_{\text{eff}}=$ constant . An increasing coupling strength $G_{\text{eff}}$ leads to an entropy increment, while more interestingly, a decreasing $G_{\text{eff}}$ would produce negative entropy for the universe.  Yet Eq.(\ref{entropy production}) only reflects the entropy production $d_pS$ on the horizon, and the total entropy change of the horizon as well as the entire universe needs further clarification within the generalized second law of thermodynamics within modified gravity. This problem is not tackled in this paper as we concentrate on the (unified) first law of thermodynamics. In addition, note that the dynamics of  $G_{\text{eff}}$ is different from the idea of varying gravitational constant in Dirac's ``large numbers hypothesis'' \cite{Dirac large numbers hypothesis}, which means nonconstancy of Newton's constant $G$ over the cosmic time scale within GR.

If we take advantage of the on-horizon dissipation equation (\ref{horizon dissipation replacement}) in $dM\,\hat{=}\,dE$, that is to say, with the assistance of the first method in Sec.~\ref{Method 1 Unified first law}, the entropy production equation (\ref{entropy production}) can be much simplified into
\begin{equation}\label{entropy production I}
T d_p S\;\hat{=}\; \Upsilon_{\text{A}}\,\frac{\dot{G}_{\text{eff}}}{G_{\text{eff}}^2}\, dt\qquad\text{and}\qquad
d_p {S}\;\hat{=}\;2\pi \Upsilon_{\text{A}}^2\frac{\dot{G}_{\text{eff}}}{G_{\text{eff}}^2} \, dt\;.
\end{equation}
It can reduce the calculations in specifying the amount of entropy production, when we need not distinguish the two sources represented by the two terms in Eq.(\ref{entropy production}). This simplification also indicates the $dM=dE$ method nicely complements the Clausius method.


\section{Further discussion on the unified formulation}\label{Discussion on the unified first law of thermodynamics}

So far a unified formulation has been developed to derive the Friedmann equations from nonequilibrium thermodynamics within generic metric gravity $R_{\mu\nu}-Rg_{\mu\nu}/2=8\pi G_{\text{eff}}  T_{\mu\nu}^{\text{(eff)}}$, and the whole operation is:

\begin{enumerate}
  \item[(1)] Inside the apparent horizon $\Upsilon<\Upsilon_{\text{A}}$, the total derivative $dM$ of the geometric mass  and the unified first law of nonequilibrium thermodynamics $dE=A\bm\psi+WdV+\mathcal{E}$ yield Friedmann equations via $dE=dM$. This method also applies to  the horizon by taking the smooth limit $\Upsilon\to\Upsilon_{\text{A}}$.
  \item[(2)] Alternatively, consider the change of total internal energy during the time interval $dt$. When evaluated on the horizon $\Upsilon=\Upsilon_{\text{A}}$,
    the extended nonequilibrium Clausius relation $TdS+Td_pS\,\hat{=}\,\delta Q$ yields the second Friedmann equation,
    which  can reproduce the first one with the continuity equation.
       \item[(3)] Derivations for the interior $\Upsilon<\Upsilon_{\text{A}}$ and the horizon $\Upsilon_{\text{A}}$  should be consistent, which sets up the thermodynamic  sign convention $T(dS+d_p S)\,\hat{=}\,\delta Q\,\hat{=}-dE\,\hat{=}-(A_{\text{A}}\bm\psi_t+\mathcal{E}_{\text{A}})$\,.
\end{enumerate}

In this section we will further investigate some problems involved in the unified formulation.

\subsection{A viability test of the extended Hawking and Misner-Sharp masses}\label{A viability test of the extended Hawking and Misner Sharp masses}

We have replaced $G$ with $G_{\text{eff}}$ to generalize the Hawking mass and the Misner-Sharp mass into Eqs.(\ref{Hawking mass}) and (\ref{Misner-Sharp mass}), respectively.  Such geometric mass worked well in deriving the Friedmann equations in the unified formulation for the correctness of this extension. Here we provide another piece of evidence by demonstrating that equality between the physical effective mass $\mathcal{M}=\rho_{\text{eff}}  V$ and the generalized geometric masses automatically reproduces the Friedmann equations.

The total derivative of the physically defined  effective mass $\mathcal{M}=\rho_{\text{eff}}V=\big(\rho_m +\rho_{\text{(MG)}}\big)\,V$ reads
\begin{equation}\label{dM test}
\begin{split}
&d\mathcal{M}\,=\,d\,\Big(\rho_{\text{eff}}V\Big)\,=\,\rho_{\text{eff}}\, dV+V\,\dot{\rho}_{\text{eff}}\,dt\\
=\;\;&\rho_{\text{eff}} A\,d\Upsilon-V\,\bigg(3H\big(\rho_{\text{eff}}+P_{\text{eff}}\big) +
\frac{\dot{G}_{\text{eff}}}{G_{\text{eff}}} \, \rho_{\text{eff}}\bigg)\, dt\\
=\;\;&4\pi \Upsilon^2 \rho_{\text{eff}}\, d\Upsilon-4\pi\Upsilon^3 H\,\Big(\rho_{\text{eff}}+P_{\text{eff}}\Big)-\frac{4}{3}\pi\, \Upsilon^3\frac{\dot{G}_{\text{eff}}}{G_{\text{eff}}} \, \rho_{\text{eff}}\, dt\;,
\end{split}
\end{equation}
where we have used the continuity equation (\ref{Generalized continuity eqn}) to replace $\dot{\rho}_{\text{eff}}$. Compare Eq.(\ref{dM test}) with Eq.(\ref{dM inside t Upsilon}),
\begin{eqnarray*}
dM\,&=&\frac{ \Upsilon^3 H}{G_{\text{eff}}}\,\left(\dot{H}-\frac{k}{a^2}   \right)\,dt
+\frac{3\Upsilon^2}{2G_{\text{eff}}}\,\left(H^2+\frac{k}{a^2}   \right)\,d\Upsilon-
\frac{\Upsilon^3 \dot{G}_{\text{eff}}}{2G_{\text{eff}}^2}\left( H^2+\frac{k}{a^2}\right)\,dt\;,
\end{eqnarray*}
and  straightforwardly, by assuming the physically defined  effective mass $\mathcal{M}=\rho_{\text{eff}}V$ equal to the geometric  effective mass in Eq.(\ref{mass}), which comes from Eqs.(\ref{Hawking mass}) and (\ref{Misner-Sharp mass}) that are defined solely out of the spacetime metric, we will automatically recover the two Friedmann equations from $d\mathcal{M}=dM$:
\begin{equation*}
H^2+\frac{k}{a^2}\,=\, \frac{8\pi G_{\text{eff}}}{3}\,\rho_{\text{eff}}\quad,\quad
\dot H-\frac{k}{a^2}\,=-4\pi G_{\text{eff}}\,\Big(\rho_{\text{eff}}+P_{\text{eff}}\Big) \;.
\end{equation*}
In this sense we argue that the generalized definitions in Eqs.(\ref{Hawking mass}) and (\ref{Misner-Sharp mass}) for the Hawking and the Misner-Sharp masses are intuitive.
Also, the equality to $\mathcal{M}=\rho_{\text{eff}}V$ indicates that
Eqs.(\ref{Hawking mass}) and (\ref{Misner-Sharp mass}) only refer to the effective matter content and do not include the  energy of  gravitational field.

Having obtained the first Friedmann equation (\ref{Friedmann eqn 1st}), we can now combine Eqs.(\ref{mass}) and  (\ref{Friedmann eqn 1st}) to eventually see that
\begin{equation}\label{mass geometric equal physical}
M_{\text{MS}}\,=\,\frac{\Upsilon^3}{2G_{\text{eff}}}\left( H^2+\frac{k}{a^2}\right)
\,=\,\frac{\Upsilon^3}{2G_{\text{eff}}}\cdot\frac{8\pi G_{\text{eff}}}{3}\,\rho_{\text{eff}}
\,=\,\frac{4}{3}\pi\Upsilon^3\,\rho_{\text{eff}}\,=\,V\,\rho_{\text{eff}}\,=\,\mathcal{M}\;,
\end{equation}
so the geometric  effective mass  Eq.(\ref{mass}) is really equal to the physically defined  mass $V\,\rho_{\text{eff}}$ with the effective density determined by Eqs.(\ref{FieldEqnGRForm}) and (\ref{Effective SEM Perfect fluid}). Note that \cite{Misner-Sharp mass III} has generalized the Misner-Sharp masses for the $f(R)$ gravity with $G_{\text{eff}}=G/f_R$ and the scalar-tensor gravity  with $G_{\text{eff}}=G/f(\phiup)$, and their results  actually
refer to the pure mass $V\rho_{m}$ of the physical matter content  compared with our generalizations, as will be clearly shown in Sec.~\ref{fR gravity} and Sec.~\ref{Scalar tensor chameleon gravity} later. Also, the following masslike function was assumed in \cite{mass-like function}
\begin{equation}
\text{Masslike}\;\coloneqq \;\frac{\Upsilon }{2G_{\text{eff}}} \,\Big( 1+ h^{\alpha\beta}\partial_\alpha \Upsilon\partial_\beta \Upsilon\Big)
\;\equiv\;\frac{\Upsilon}{2G_{\text{eff}}}\,\bigg(2-\frac{\Upsilon^2}{\Upsilon^2_{\text{A}}}\bigg)
\;\;\hat{=}\;\;\frac{\Upsilon_{\text{A}}}{2G_{\text{eff}}}\;,
\end{equation}
in an attempt to recover the Friedmann equations on the horizon itself from the equilibrium Clausius relation without the entropy-production  correction $d_pS$. However, it is not suitable in our more general formulation in Sec.~\ref{Inside the cosmological apparent horizon} and Sec.~\ref{On the cosmological apparent horizon}, especially in the $dM=dE$ approach for the whole region $\Upsilon\leq \Upsilon_{\text{A}}$, and it does not pass the test just above as in Eq.(\ref{dM test}).

On the other hand, recall that in recent studies on  the interesting idea of ``chemistry'' of anti-de Sitter black holes \cite{Chemistry of AdS black holes}, the mass $M$ has
been treated as the enthalpy $\mathcal{H}$ rather than total internal energy $E$, i.e. $M=\mathcal{H}=E+PV$
where the pressure $P$ is proportional to the cosmological constant $\Lambda$. Since $\Lambda$ the the simplest modified-gravity term, similarly, is it possible
to identify the mass $M$ in a sphere of radius $\Upsilon\leq\Upsilon_{\text{A}}$ in the FRW universe as the enthalpy $\mathcal{H}=E+\widetilde{P}V$ for some kind of pressure $\widetilde{P}$ (it can be $P_{\text{eff}}$,  $P_{\text{(MG)}}$, etc.)? We find that
the answer seems to be negative. The equality
between Eqs.(\ref{dM inside t r})(\ref{dM inside t Upsilon}) for $dM$ and Eqs.(\ref{dE inside t r})(\ref{dE inside t Upsilon}) for $dE$,
as well as the consistency among Eqs.(\ref{dM inside t Upsilon}), (\ref{dE inside t Upsilon}) and (\ref{dM test}) clearly shows that the mass $M$ should be identified as the total internal energy $E$. Moreover, if forcing the equality $M=\mathcal{H}$, then $dM=d\mathcal{H}=d(E+\widetilde{P}V)$ implies that necessarily that $\widetilde{P}\equiv0$ and $\dot{\widetilde{P}}\equiv0$ and thus we still have $M\equiv E$.


\subsection{The continuity/conservation equation}\label{The continuity conservation equation}

As emphasized before in Sec.~\ref{Introduction}, we are considering ordinary modified gravity under minimal geometry-matter coupling, $\mathscr{L}_{\text{total}}=\mathscr{L}_{\text{gravity}}+16\pi G\mathscr{L}_m$, with an isolated matter density $\mathscr{L}_m$ in the total lagrangian density and thus no curvature-matter coupling terms like $R\mathscr{L}_m$; or equivalently, the gravity/geometry part and the matter part in the total action are fully separable,  $\mathcal{I}_{\text{total}}=\mathcal{I}_{\text{gravity}}+\mathcal{I}_m$. For the matter action $\mathcal{I}_m=\int d^4x \sqrt{-g}\mathscr{L}_m$ itself, the SEM tensor $T^{(m)}_{\mu\nu}$ is defined by the following stationary variation (eg. \cite{AA Tian-Booth Paper}),
\begin{equation}\label{continuity conservation I}
\delta \mathcal{I}_m\,=\,\delta \int d^4x \sqrt{-g}\,\mathscr{L}_m\, =-\frac{1}{2}\int  d^4x \sqrt{-g}\,T_{\mu\nu}^{(m)}\,\delta g^{\mu\nu}
\quad\text{ with }\quad
T_{\mu\nu}^{(m)}\,\coloneqq\, \frac{-2}{\sqrt{-g}} \,\frac{\delta\, \Big(\!\!\sqrt{-g}\,\mathscr{L}_m \Big)}{\delta g^{\mu\nu}}\;.
\end{equation}
On the other hand, since $\mathscr{L}_m$ is a scalar invariant, Noether's conservation law yields
\begin{equation}\label{continuity conservation II}
\nabla^\mu \,\left( \frac{1}{\sqrt{-g}} \,\frac{\delta\, \Big(\!\!\sqrt{-g}\,\mathscr{L}_m \Big)}{\delta g^{\mu\nu}}\right)=\,0\;.
\end{equation}
Comparison with Eq.(\ref{continuity conservation I}) yields that Eq.(\ref{continuity conservation II}) can be rewritten into $\displaystyle -\frac{1}{2}\nabla^\mu T_{\mu\nu}^{(m)}=0$. Hence, the definition of  the SEM tensor $T^{(m)}_{\mu\nu}$ as in Eq.(\ref{continuity conservation I}) is Noether-compatible, and the definition of $T^{(m)}_{\mu\nu}$ by itself automatically guarantees stress-energy-momentum conservation
\begin{equation}
\nabla^\mu T_{\mu\nu}^{(m)}\,=\,0\;.
\end{equation}
For a time-dependent perfect-fluid matter content $T^{\mu\,\text{(m)}}_{\;\;\nu}=\text{diag}\,[-\rho_m(t)\,, P_m(t)\,, P_m(t)\,, P_m(t)]$ (say for the FRW universe), $\nabla^\mu T_{\mu\nu}^{(m)}=0$ gives rise to the continuity equation
\begin{equation}\label{Matter continuity Noether}
\dot{\rho}_m+3H\,\Big(\rho_m+P_m  \Big)\,=\,0\;.
\end{equation}
Hence, the total continuity equation (\ref{Generalized continuity eqn}) can be reduced into
\begin{equation}
\dot{\rho}_{\text{(MG)}}+3H\,\Big(\rho_{\text{(MG)}}+P_{\text{(MG)}}  \Big)
\,=\,-\frac{\dot{G}_{\text{eff}} }{G_{\text{eff}} }\,\Big(\rho_m+\rho_{\text{(MG)}}  \Big)\;.
\end{equation}
Also, note that $\rho_m$ collects the energy density of all possible physical material content,
\begin{equation}
\rho_m\,=\,\sum\rho_{m(i)}\,=\,\rho_m(\text{baryon dust})+\rho_m(\text{radiation})+\rho_m(\text{dark energy})+\rho_m(\text{dark matter})+\cdots\;,
\end{equation}
and for each type of  component $\rho_{m(i)}$, by decomposing Eq.(\ref{Matter continuity Noether}) we have individually
\begin{equation}
\dot{\rho}_{m(i)}+3H\,\Big(\rho_{m(i)}+P_{m(i)}  \Big)\,=\,Q_{m(i)}\quad \text{with} \quad
\sum Q_{m(i)}\,=\,0\;,
\end{equation}
where $Q_{m(i)}$ denotes the  energy exchange due to the possible self- and cross-interactions among different matter components.

These results are applicable to the situation of minimal geometry-matter couplings. The thermodynamics of nonminimally coupled theories like  $\mathscr{L}=f(R\,,T^{(m)})+16\pi G \mathscr{L}_m$ \cite{Nonminimal coupling fRT} (where $T^{(m)}=g_{\mu\nu}T^{\mu\nu}_{(m)}$) and  $\mathscr{L}=f(R\,,T^{(m)}\,,R_{\mu\nu}T^{\mu\nu}_{(m)})+16\pi G \mathscr{L}_m$ \cite{Nonminimal coupling fRTmunu} have been attempted using the traditional formulation as in \cite{Cai II} for $f(R)$ gravity.  However, more profound thermodynamic properties
may hide in these theories, as there is direct energy exchange between spacetime geometry and the energy-matter content
under nonminimal curvature-matter couplings \cite{Nonminimal coupling 0, Nonminimal coupling, AA Tian-Booth Paper}. For example,
very recently Harko \cite{Nonminimal coupling I} has interpreted the
generalized conservation equations in $\mathscr{L}=f(R\,, \mathscr{L}_m)$ and $\mathscr{L}=f(R\,,T^{(m)})+16\pi G \mathscr{L}_m$ gravity as a matter creation process with an irreversible energy flow from the gravitational field to the created matter  in accordance with the second law of thermodynamics.
The unusual thermodynamic effects in these theories go beyond the scope of this paper, but for the chameleon effect \cite{scalar tensor chameleeon, scalar tensor chameleeon II} which is another type of nonminimal coupling in scalar-tensor alternative gravity, we manage to find the extra energy dissipation and entropy production  caused by the chameleon field, as will be shown later in Sec.~\ref{Scalar tensor chameleon gravity}.


\subsection{``Negative temperature'' on the horizon could remove the entropy production $d_pS$}

In Sec.~\ref{Method 2 Clausius relation}, by studying the energy change during $dt$ across the horizon we have derived the second Friedmann equation from the nonequilibrium Clausius relation $T(dS+d_p S) \hat{=}-  \big(A_{\text{A}}\bm\psi_t+\mathcal{E}_{\text{A}}\big)$ with a necessary entropy-production element $d_pS$. However, we also observe that if the geometric temperature of the horizon were to be defined by the following ``negative temperature''
\begin{equation}\label{Neg T}
\mathcal{T}\,\equiv \,-\frac{1}{2 \pi \Upsilon_{\text{A}}}\,<0\;,
\end{equation}
which is the opposite to Eq.(\ref{T}),
then it is easily seen from Sec.~\ref{Method 2 Clausius relation} that
\begin{equation}\label{NegTdS TdpS}
\begin{split}
\mathcal{T}dS- A_{\text{A}}\bm\psi_t-\mathcal{E}_{\text{A}}
\;\hat{=} \;\left(\frac{ \dot{\Upsilon}_{\text{A}}}{G_{\text{eff}}}dt -  A_{\text{A}}\bm\psi_t\right)
-&\left(\frac{1}{2}\Upsilon_{\text{A}}\frac{\dot{G}_{\text{eff}}}{G_{\text{eff}}^2} dt+\mathcal{E}_{\text{A}}\right)\\
\hat{=} -\left(\frac{ H \Upsilon_{\text{A}}^3}{G_{\text{eff}}}
\,\Big( \dot{H}-\frac{k}{a^2}\Big)
+ A_{\text{A}}\,\big(\rho_{\text{eff}}+P_{\text{eff}}\big)\,H\Upsilon_{\text{A}}\right)\,dt
-&\left(\frac{1}{2}\Upsilon_{\text{A}}\frac{\dot{G}_{\text{eff}}}{G_{\text{eff}}^2} -\frac{4}{3}\pi\, \Upsilon_{\text{A}}^3\frac{\dot{G}_{\text{eff}}}{G_{\text{eff}}} \, \rho_{\text{eff}} \right)\,dt\,.
\end{split}
\end{equation}
In the last row of Eq.(\ref{NegTdS TdpS}), the vanishing of the former parentheses leads to the second Friedmann equation, while in the second parentheses, the $\dot{G}_{\text{eff}}$ component of $\mathcal{T}dS$ and the overall energy dissipation term $\mathcal{E}_{\text{A}}$  cancel out each other to yield the first Friedmann equation.
Hence, with the negative horizon temperature Eq.(\ref{Neg T}), both Friedmann equations could be obtained from the standard equilibrium Clausius relation
\begin{equation}\label{NegTdS nonequi}
\begin{split}
\mathcal{T}dS\; \hat{=}\;dE\;\;\hat{=}\;  A_{\text{A}}\bm\psi_t+\mathcal{E}_{\text{A}}
\end{split}
\end{equation}
without employing an entropy-production term $d_pS$.

However, the negative temperature ansatz Eq.(\ref{Neg T}) is problematic in various aspects. For example,  negative absolute temperature is forbidden by the third law of thermodynamics (as is well known, the so-called ``negative temperature'' state in atomic physics actually occurs at a unusual phase of very high temperature where the entropy decreases with increasing internal energy, $T^{-1}\coloneqq {\partial S}/{\partial E}<0$).
Also,  if tracing back to the past history of the expanding Universe, one will find
the horizon carrying a more and more negative temperature $\mathcal{T}$ while enclosing a more and more (positively) hot interior.  From these perspectives, the observation from Eq.(\ref{Neg T})  that $\mathcal{T}=-1/(2 \pi \Upsilon_{\text{A}})$ could provide a most economical way to recover the Friedmann equations on the apparent horizon from equilibrium thermodynamics may just be an interesting coincidence.


\subsection{Equilibrium situations with $G_{\text{eff}}=G=\text{constant}$  and thus $\mathcal{E}=0$}\label{Situations of G without dissipative energy}

When the effective gravitational coupling strength $G_{\text{eff}}$ reduces to become Newton's constant $G$,
the field equation (\ref{FieldEqnGRForm}) reduces to
\begin{equation}
R_{\mu\nu}-\frac{1}{2}Rg_{\mu\nu} \,=\, 8\pi G  \,T_{\mu\nu}^{\text{(eff)}}
\,=\,  8\pi G  \,\Big(T_{\mu\nu}^{(m)}+T_{\mu\nu}^{\text{(MG)}}\Big)\;.
\end{equation}
For theories in this situation, the Lagrangian density generally takes the form
\begin{equation}\label{action G}
\mathscr{L}\,=\,R+
f(R_{\mu\nu}R^{\mu\nu}\,,R_{\mu\alpha\nu\beta}R^{\mu\alpha\nu\beta}\,,\mathcal{R}_{\,i}\cdots )+
\omegaup\,\big(\phiup\,, \nabla_\mu\phiup\nabla^\mu\phiup\big) +16\pi G\mathscr{L}_m\;,
\end{equation}
where $\mathcal{R}_i$ denotes an arbitrary algebraic or differential
Riemannian invariant $\mathcal{R}_i=\mathcal{R}_i\,\big(g_{\alpha\beta}\,,R_{\mu\alpha\nu\beta}\,,\nabla_\gamma R_{\mu\alpha\nu\beta}\,,\ldots\,,$ $
\nabla_{\gamma_ 1}\!\nabla_{\gamma_ 2}\ldots\nabla_{\gamma_ q} R_{\mu\alpha\nu\beta}\big)$ which is beyond the Ricci scalar $R$ and makes no contribution to the coefficient of $R_{\mu\nu}$ in the field equation. $\omegaup$ is a generic function of the scalar field $\phiup=\phiup(x^\mu)$ and its kinetic term $ \nabla_\mu\phiup\nabla^\mu\phiup$. For example, the $\mathscr{L}=R+f(R_{\mu\nu} R^{\mu\nu}\,,R_{\mu\alpha\nu\beta}R^{\mu\alpha\nu\beta})+16\pi G\mathscr{L}_m$ fourth-order gravity and typical scalarial dark-energy models \cite{Dark Energy} (like quintessence, phantom, k-essence)
all belong to this class.

To apply the unified formulation developed in
Sec.~\ref{Inside the cosmological apparent horizon} and
Sec.~\ref{On the cosmological apparent horizon} for this situation,
we just need to replace $G_{\text{eff}}$ by $G$, set $\dot{G}_{\text{eff}}=0$,
and remove the energy dissipation term $\mathcal{E}$. Hence, the Hawking or Misner-Sharp
mass enclosed by a sphere of radius $\Upsilon$ is
$M=(\Upsilon^3/2G)\big( H^2+k/{a^2}\big)$.
Compare the transverse gradient $dM$ of the mass with the change of internal energy $dE=A\bm \psi+WdV$, and by matching the coefficients of
\begin{equation}\label{G dM dE r}
\begin{split}
dM\,&=\,\frac{ \Upsilon^3 H}{2G}\,\left(2\dot{H}+3H^2+\frac{k}{a^2}   \right)\,dt
+\frac{3\Upsilon^2}{2G}\,\left(H^2+\frac{k}{a^2}   \right)\,adr\\
dE\,&=\,-4\pi \Upsilon^3 \,H \,P_{\text{eff}}\,dt+4\pi\Upsilon^2\,\rho_{\text{eff}}\, adr
\end{split}
\end{equation}
in the comoving coordinates $(t\,,r)$\,, or
\begin{equation}\label{G dM dE Upsilon}
\begin{split}
dM\,&=\,\frac{ \Upsilon^3 H}{G}\,\left(\dot{H}-\frac{k}{a^2}   \right)\,dt
+\frac{3\Upsilon^2}{2G}\,\left(H^2+\frac{k}{a^2}   \right)\,d\Upsilon\\
dE\,&=\,-4\pi \Upsilon^3 \,H \,\Big(\rho_{\text{eff}}+P_{\text{eff}}\Big)\,dt+4\pi \Upsilon^2\,\rho_{\text{eff}}\,d\Upsilon\;,
\end{split}
\end{equation}
in the astrophysical areal coordinates $(t\,,\Upsilon)$,
one obtains the Friedmann equations with $G_{\text{eff}}=G$:
\begin{equation}
H^2+\frac{k}{a^2}\,=\, \frac{8\pi G }{3}\,\rho_{\text{eff}}\quad\text{and}\quad
\dot H-\frac{k}{a^2}\,=-4\pi G\,\Big(\rho_{\text{eff}}+P_{\text{eff}}\Big)\quad\text{or}\quad
2\dot H+3H^2+\frac{k}{a^2}\,= -8\pi G  P_{\text{eff}}\;.
\end{equation}
Moreover, in the smooth limit $\Upsilon\to \Upsilon_{\text{A}}$ Eqs. (\ref{G dM dE r}) and (\ref{G dM dE Upsilon}) recover the complete set of Friedmann equations on the apparent horizon $\Upsilon=\Upsilon_{\text{A}}$ by $dM\,\hat{=}\,dE$.
Alternatively, with the absolute temperature $T$ and the entropy $S$ of the horizon being
\begin{equation}\label{T S G}
T\;\hat{=}\;\frac{1}{2 \pi \Upsilon_{\text{A}}} \qquad\text{and}\qquad
S\;\hat{=}\;\frac{A_{\text{A}}}{4G}\;\hat{=}\;\frac{\pi \Upsilon_{\text{A}}}{G} \;,
\end{equation}
we have
\begin{equation}\label{G TdS horizon}
TdS\,=\,  \frac{ \dot{\Upsilon}_{\text{A}}}{G}\,dt
\quad\text{and}\quad A_{\text{A}}\bm\psi_t\;\hat{=}\; -A_{\text{A}}\,\Big(\rho_{\text{eff}}+P_{\text{eff}}\Big)\,H\Upsilon_{\text{A}}\,dt\;.
\end{equation}
Thus, the equilibrium Clausius relation $TdS\,\hat{=}\,\delta Q\,\hat{=}-A_{\text{A}}\bm\psi_t$ with Eq.(\ref{G TdS horizon})  for an isochoric process leads to the second Friedmann equation
$
\dot H-k/{a^2}\,\hat{=}-4\pi G\big(\rho_{\text{eff}}+P_{\text{eff}}\big)\;.
$
Taking into account the continuity equation with vanishing dissipation $\mathcal{E}=0$:
\begin{equation}\label{G continuity eqn}
\dot{\rho}_{\text{eff}}+3H\,\big(\rho_{\text{eff}}+P_{\text{eff}}  \big)\,=\,0\;,
\end{equation}
integration of the second Friedmann equation leads to the first equation
$
H^2+k/{a^2} = 8\pi G\,\rho_{\text{eff}}/3,
$
where the integration constant has been neglected or absorbed into $\rho_{\text{eff}}$.
Moreover, the continuity/conservation equation (\ref{G continuity eqn}) together with conservation of $T_{\mu\nu}^{(m)}$ in Eq.(\ref{Matter continuity Noether}) lead to
\begin{equation}
\begin{split}
\dot{\rho}_{\text{(MG)}}+3H\,\Big(\rho_{\text{(MG)}}+P_{\text{(MG)}}  \Big)\,=\,0\;.
\end{split}
\end{equation}
For the componential convariant Lagrangian density $\sqrt{-g}\,f(R_{\mu\nu}R^{\mu\nu}\,,R_{\mu\alpha\nu\beta}R^{\mu\alpha\nu\beta}\,,\mathcal{R}_{\,i}\cdots )$ in Eq.(\ref{action G}), this is actually the ``generalized contracted Bianchi identities''  \cite{AA Tian-Booth Paper} in perfect-fluid form under the FRW background.


\section{Examples}\label{Examples}

In this section, we will apply the unified formulation  in Sec.~\ref{Inside the cosmological apparent horizon} and Sec.~\ref{On the cosmological apparent horizon} to some concrete theories of modified gravity.
Compatible with the FRW metric Eq.(\ref{FRW metric I}) in the signature $(-,+++)$,
we will adopt the geometric sign convention  $\Gamma^\alpha_{\delta\beta}=\Gamma^\alpha_{\;\;\,\delta\beta}$ ,  $R^{\alpha}_{\;\;\beta\gamma\delta}=\partial_\gamma \Gamma^\alpha_{\delta\beta}-\partial_\delta \Gamma^\alpha_{\gamma\beta}\cdots$ and  $R_{\mu\nu}=R^\alpha_{\;\;\mu\alpha\nu}$.


\subsection{$f(R)$ gravity}\label{fR gravity}

The $f(R)$ gravity \cite{Example fR} is the simplest class of fourth-order gravity, which straightforwardly
 generalizes the Hilbert-Einstein Lagrangian density $\mathscr{L}_{\text{HE}}=R+16\pi G \mathscr{L}_m$ into $\mathscr{L}=f(R)+16\pi G \mathscr{L}_m$ by replacing the Ricci scalar $R$ with its arbitrary function $f(R)$.
The field equation in the form of Eq.(\ref{FieldEqnGRForm}) is
\begin{equation}\label{Field eqn fR}
\begin{split}
R_{\mu\nu}-\frac{1}{2}\,Rg_{\mu\nu}\,&=\, 8\pi\frac{ G}{f_R} T_{\mu\nu}^{(m)}+  \frac{1}{f_R}\,\Bigg(
\,\frac{1}{2}\big(f-f_R R\big)\,g_{\mu\nu}+\big(\nabla_\mu\!\nabla_\nu-g_{\mu\nu}\Box\big)\,f_R  \Bigg)\;,\\
\end{split}
\end{equation}
where $f_R\coloneqq \partial f(R)/\partial R$ and $\Box\equiv \nabla^\alpha \nabla_\alpha$ denotes the covariant d'Alembertian.
From the coefficient of $T_{\mu\nu}^{(m)}$ we learn that the effective gravitational coupling strength for $f(R)$ gravity is 
\begin{equation}
G_{\text{eff}}\,=\,\frac{G}{f_R}\;,
\end{equation}
and thus the modified-gravity SEM tensor is
\begin{equation}
T_{\mu\nu}^{\text{(MG)}}\,=\,\frac{1}{8\pi G}\,\Bigg(\frac{1}{2}\big(f-f_R R\big)g_{\mu\nu}
+\big(\nabla_\mu\!\nabla_\nu-g_{\mu\nu}\Box\big)f_R  \Bigg)\;,
\end{equation} which has collected the contributions from nonlinear and fourth-order curvature terms. Substituting the FRW metric Eq.(\ref{Friedmann eqn 1st}) into this $T_{\mu\nu}^{\text{(MG)}}$ and keeping in mind $T^{\mu\,\text{(MG)}}_{\;\;\nu}=$ $\text{diag}[-\rho_{\text{(MG)}},P_{\text{(MG)}},
P_{\text{(MG)}},P_{\text{(MG)}}]$, the energy density and pressure from the $f(R)$ modified-gravity effect are found to be
\begin{equation}\label{fR rho P}
\rho_{\text{(MG)}}\,=\; \frac{1}{8\pi G}\, \Big(\frac{1}{2}f_R R-\frac{1}{2}f-3H\,\dot{f}_R\,\Big) \qquad\text{and}\qquad
P_{\text{(MG)}}\, =\; \frac{1}{8\pi G}\,\Big(\frac{1}{2}f-\frac{1}{2}f_R R+\ddot{f}_{R}+2H\,\dot{f}_R \Big)\;.
\end{equation}
Given $G_{\text{eff}}=G/f_R$, the Hawking or Misner-Sharp mass in a sphere of radius $\Upsilon$ in the universe is
\begin{equation}\label{fR mass}
M\,=\,\frac{f_R\Upsilon^3}{2G}\left( H^2+\frac{k}{a^2}\right)\qquad\text{with}\qquad
M_{\text A}\;\hat{=}\;\frac{f_R\Upsilon_{\text A}}{2G}\;.
\end{equation}
Also, the geometric nonequilibrium energy dissipation term associated with  $G_{\text{eff}}$ and  the geometric Wald-Kodama entropy of the horizon $\Upsilon_{\text{A}}$
\begin{equation}\label{fR E S}
\mathcal {E}\,= \,\frac{4}{3}\pi\, \Upsilon^3\frac{\dot{f}_R}{f_R} \, \rho_{\text{eff}}\; dt  \qquad\text{and}\qquad
S\,=\,\frac{A_{\text A}f_R}{4G}\;.
\end{equation}
Note that in $\mathcal{E}$ the term $\frac{4}{3}\pi\, \Upsilon^3\, \rho_{\text{eff}}$ should not be combined into the mass $V\rho_{\text{eff}}=\mathcal{M}$ at this stage for the reason stressed after Eq.(\ref{Dissipation energy density}). Applying the unified formulation developed in Sec.~\ref{Inside the cosmological apparent horizon} and Sec.~\ref{On the cosmological apparent horizon} to the FRW universe governed by $f(R)$ gravity, for the interior and the horizon $\Upsilon\leq \Upsilon_{\text A}$, the unified first law $dE=A\bm\psi+WdV+\mathcal{E}=dM$ of nonequilibrium thermodynamics and the nonequilibrium Clausius relation $T(dS+d_PS)\,\hat{=}\,\delta Q\,\hat{=}-(A_{\text A}\bm\psi+\mathcal{E}_{\text A})$ give rise to
\begin{equation}\label{fR Friedmann eqn 1}
H^2+\frac{k}{a^2}  \,=\,\frac{8\pi}{3}\,\frac{G}{f_R}\,\rho_m +\frac{1}{3 f_R}\, \Big(\frac{1}{2}f_R R-\frac{1}{2}f-3H\,\dot{f}_R\,\Big) \;,
\end{equation}
\begin{equation}\label{fR Friedmann eqn 2}
\hspace{0mm}\dot H-\frac{k}{a^2}\,= -4\pi \,\frac{G}{f_R}\, \Big(\rho_m+P_m\Big)
- \frac{1}{2f_R}\,\Big(\ddot{f}_{R}-H\,\dot{f}_R \Big)\;.
\end{equation}
In the meantime, the nonequilibrium entropy production $d_pS$ on the horizon turns out to be
\begin{equation}\label{fR dpS}
d_p {S}\;\hat{=}\;-2\pi \Upsilon_{\text{A}}^2\frac{\dot{f}_R}{G} \, dt\,\;.
\end{equation}

Substituting the FRW metric Eq.(\ref{FRW metric I}) into Eq.(\ref{Field eqn fR}), we have verified that, Eqs.(\ref{fR Friedmann eqn 1}) and (\ref{fR Friedmann eqn 2}) are exactly the Friedmann equations of the FRW universe in $f(R)$ gravity. Such thermodynamics-gravity correspondence within $f(R)$ gravity has been investigated before in \cite{Cai II, Eling Nonequilibrium Thermodynamics} with different setups for the quantities $\{M\,,\rho_{\text{(MG)}}\,,P_{\text{(MG)}}\cdots\}$ and thus $\{\bm\psi\,,W\cdots\}$; compared with these earlier works, we have revised the thermodynamic setups and improved the result of entropy production.

Also note that, compact notations have been used in Eqs.(\ref{fR Friedmann eqn 1}) and (\ref{fR Friedmann eqn 2}), and $f_R$ itself is treated as a function of the comoving time $t$. Otherwise, one can further write $\dot{f}_R$ into $f_{RR}\,\dot R$ and $\ddot{f}_R$ into $f_{RR}\,\dot R+ f_{RRR}\,\dot{R}^2$ as in \cite{Cai II, Cai V}, and for the FRW spacetime with metric Eq.(\ref{FRW metric I}), we have already known the Ricci scalar that
\begin{equation}
R\,=\,R(t)\,=\,6\,\Big(\dot H+2H^2+\frac{k}{a^2}\Big)\;,
\end{equation}
which in turn indicates the third-derivative $\dddot H$ and thus fourth-derivative $\ddddot a$ get involved in Eqs.(\ref{fR Friedmann eqn 1}) and (\ref{fR Friedmann eqn 2}), and these terms are gone once we return to GR with $f_R=1$.

In \cite{Misner-Sharp mass III}, Cai et al. have generalized the Misner-Sharp energy/(mass) to $f(R)$ gravity by the integration and the conserved-charge methods. Specifically for the FRW universe, they found that the energy/mass within a sphere of radius $\Upsilon$ is
\begin{equation}\label{Cai MS mass fR}
\begin{split}
E_{\text{eff}}\,&=\,\frac{\Upsilon}{2G}\,\Bigg(\big(1-h^{\alpha\beta}
\partial_\alpha \Upsilon\partial_\beta \Upsilon\big)+\frac{1}{6}\,\Upsilon^2\,\big(f-f_R R \big)
-\Upsilon \,h^{\alpha\beta} \,\partial_\alpha f_R \,\partial_\beta \Upsilon\,\Bigg)\\
&=\,\frac{\Upsilon^3}{2G}\,\bigg( \frac{1}{\Upsilon_{\text{A}}^2}\,f_R+
\frac{1}{6}\,\big(f-f_R R \big)+H\,\dot{f}_R \bigg)\;,
\end{split}
\end{equation}
with $\Upsilon_{\text{A}}=1/\sqrt{H^2+k/a^2}$\,. What are the differences between this $E_{\text{eff}}$ and our extended Misner-Sharp mass in Eqs.(\ref{Misner-Sharp mass}) and (\ref{mass})
in this paper? In the first and second row of Eq.(\ref{Cai MS mass fR}), the first terms therein are respectively the definition Eq.(\ref{Misner-Sharp mass}) and the concrete mass Eq.(\ref{mass}) in our usage. To further understand the remaining terms in Eq.(\ref{Cai MS mass fR}), one can manipulate it into
\begin{equation}\label{Cai MS mass fR II}
\begin{split}
E_{\text{eff}}\,
\,&=\,\frac{f_R\Upsilon^3}{2G}\Big( H^2+\frac{k}{a^2}\Big)-\frac{\Upsilon^3}{2G}\,
\bigg(\frac{1}{6}\big(f_R R-f \big)-H\dot{f}_R \bigg)\\
&=\,\frac{f_R\Upsilon^3}{2G}\Big( H^2+\frac{k}{a^2}\Big)-\frac{4}{3}\pi\Upsilon^3\cdot\frac{1}{8\pi G}\,
\Big(\frac{1}{2}f_R R-\frac{1}{2}f-3H\,\dot{f}_R\, \Big)\;.
\end{split}
\end{equation}
Recall that in Eq.(\ref{mass geometric equal physical}), we have already proved the geometric mass Eq.(\ref{mass}) with which we start our formulation is equal to the physically defined mass  $\rho_{\text{eff}}\,V=\big(\rho_{m}+\rho_{\text{(MG)}}\big)\,V$\,.
Then from the density $\rho_{\text{(MG)}}$ in Eq.(\ref{fR rho P}) and the mass $M$ in Eq.(\ref{fR mass}) for $f(R)$ gravity in our unified formulation, it turns out that the $E_{\text{eff}}$ in Eq.(\ref{Cai MS mass fR II}) is actually
\begin{equation}\label{Cai MS mass fR III}
\begin{split}
E_{\text{eff}}\,=\,M-\rho_{\text{(MG)}}V\,=\,\Big(\rho_{m}+\rho_{\text{(MG)}}\Big)\,V-\rho_{\text{(MG)}}V
\,=\,\rho_{m} V\;.
\end{split}
\end{equation}
Hence, the ``generalized Misner-Sharp energy $E_{\text{eff}}$'' in \cite{Misner-Sharp mass III} for the FRW universe within $f(R)$ gravity
exactly match the pure mass of the physical matter content in our formulation of $f(R)$ cosmology.


\subsection{Generalized Brans-Dicke gravity with self-interaction potential}\label{Generalized Brans Dicke gravity with self-interaction potential}

Now, consider a generalized Brans-Dicke gravity with self-interaction potential in the Jordan frame given by the following Lagrangian density,
\begin{equation}\label{Generalized Brans-Dicke with self-interaction potential}
\mathscr{L}_{\text{GBD}}\,=\, \phiup R-\frac{\omega(\phiup)}{\phiup} \,\nabla_{\alpha}\phiup \nabla^{\alpha}\phiup-V(\phiup)+16\pi G\mathscr{L}
_m \;,
\end{equation}
where, to facilitate the comparison with the proceeding
case of $f(R)$ gravity, we have adopted the convention with an explicit $G$ in  $16\pi G\mathscr{L}
_m$\,, rather than just $16\pi \mathscr{L}_m$ which encodes $G$ into $\phiup^{-1}$ \cite{Brans Dicke}.
The gravitational field equation $\delta (\!\sqrt{-g}\,\mathscr{L}_{\text{GBD}})/\delta g^{\mu\nu}=0$ is
\begin{equation}\label{Generalized Brans-Dicke 1}
R_{\mu\nu}-\frac{1}{2}R g_{\mu\nu}\,=\,8\pi \frac{G}{\phiup} T^{(m)}_{\mu\nu}
+\frac{1}{\phiup}\,\big(\nabla_\mu\nabla_\nu-g_{\mu\nu}\Box\big)\,\phiup+\frac{\omega(\phiup)}{\phiup^2}\,\Big( \nabla_{\mu} \phiup  \nabla_{\nu}\phiup-\frac{1}{2}g_{\mu\nu} \,\nabla_{\alpha}\phiup \nabla^{\alpha}\phiup \Big)
-\frac{1}{2\phiup}V\,g_{\mu\nu}\;,
\end{equation}
from which we directly read that the effective coupling strength and the modified-gravity SEM tensor are
\begin{equation}
G_{\text{eff}}\,=\,\frac{G}{\phiup}
\qquad\mbox{and}
\end{equation}
\begin{equation}
T_{\mu\nu}^{\text{(MG)}}\,=\,\frac{1}{8\pi G}\,\Bigg(\big(\nabla_\mu\nabla_\nu-g_{\mu\nu}\Box\big)\,\phiup+\frac{\omega(\phiup)}{\phiup}\Big( \nabla_{\mu} \phiup  \nabla_{\nu}\phiup-\frac{1}{2}g_{\mu\nu} \,\nabla_{\alpha}\phiup \nabla^{\alpha}\phiup \Big)
-\frac{1}{2}V\,g_{\mu\nu} \Bigg)\;,
\end{equation}
where $T_{\mu\nu}^{\text{(MG)}}$ encodes the gravitational effects of the scalar field $\phiup$. Put the FRW metric Eq.(\ref{Friedmann eqn 1st}) back to $T_{\mu\nu}^{\text{(MG)}}$ with $T^{\mu\,\text{(MG)}}_{\;\;\nu}$$=\text{diag}[-\rho_{\text{(MG)}},P_{\text{(MG)}},
P_{\text{(MG)}},P_{\text{(MG)}}]$, and  the energy density and pressure from $\phiup$ are found to be
\begin{equation}
\rho_{\text{(MG)}}  \,=\,\frac{1}{8\pi G}\bigg(-3H\dot\phiup+\frac{\omega}{2\phiup}\dot{\phiup}^2+\frac{1}{2}V\bigg)\quad\mbox{with}\quad
P_{\text{(MG)}}\,=\,\frac{1}{8\pi G}\bigg(\ddot{\phiup}+2H\dot \phiup+\frac{\omega}{2\phiup}\dot{\phiup}^2-\frac{1}{2}V \bigg)\;.
\end{equation}
since $G_{\text{eff}}={G}/{\phiup}$, the geometric mass enveloped in a sphere of radius $\Upsilon$ is
\begin{equation}
M\,=\,\frac{\phiup\Upsilon^3}{2G}\left( H^2+\frac{k}{a^2}\right)\qquad\text{with}\qquad
M_{\text A}\;\hat{=}\;\frac{\phiup\Upsilon_{\text A}}{2G}\;,
\end{equation}
which in fact matches the Misner-Sharp mass of black holes in standard Brans-Dicke gravity in \cite{BH Mass Brans Dicke}.
Also the nonequilibrium energy dissipation term $\mathcal {E}$ associated with the evolution of $G_{\text{eff}}$ and the Wald-Kodama entropy $S$ of the horizon are
\begin{equation}
\mathcal {E}\,= \,\frac{4}{3}\pi\, \Upsilon^3\frac{\dot{\phiup}}{\phiup} \, \rho_{\text{eff}}\; dt  \qquad\text{and}\qquad
S\;\hat{=}\;\frac{A_{\text A}\phiup}{4G}\;.
\end{equation}
Following the unified formulation developed in Sec.~\ref{Inside the cosmological apparent horizon} and Sec.~\ref{On the cosmological apparent horizon} to study $dM=dE=A\bm\psi+WdV+\mathcal{E}$ for the region $\Upsilon\leq \Upsilon_{\text A}$ and $T(dS+d_PS)\,\hat{=}\,\delta Q\,\hat{=}-(A_{\text A}\bm\psi+\mathcal{E}_{\text A})$ for the horizon itself, we find
\begin{equation}\label{GBD Friedmann 1}
H^2+\frac{k}{a^2}\,=\,\frac{8\pi}{3}\frac{G}{\phiup}\,\rho_m + \frac{1}{3\phiup}\,\bigg(
-3H\dot\phiup+\frac{\omega}{2\phiup}\dot{\phiup}^2+\frac{1}{2}V\bigg)\;,
\end{equation}
\begin{equation}\label{GBD Friedmann 2}
\hspace{2mm}\dot H-\frac{k}{a^2}\,=-4\pi\frac{G}{\phiup} \,\Big(\rho_m+P_m\Big)
-\frac{1}{2\phiup}\,\bigg(\ddot{\phiup}-H\dot \phiup  +\frac{\omega}{\phiup}\dot{\phiup}^2 \bigg)\;,
\end{equation}
where as we can see, the scalar kinetics
$\frac{\omega(\phiup)}{\phiup} \,\nabla_{\alpha}\phiup \nabla^{\alpha}\phiup$ and the potential
$V(\phiup)$ does not influence the evolution of the Hubble parameter $H$,
and meanwhile the dynamics of $\phiup$ and its nonminimal coupling to $R$ in Eq.(\ref{Generalized Brans-Dicke with self-interaction potential}) leads to
the entropy production
\begin{equation}
d_p {S}\;\hat{=}\;-2\pi \Upsilon_{\text{A}}^2\frac{\dot{\phiup}}{G} \, dt
\end{equation}
for the horizon. We have already verified that Eqs.(\ref{GBD Friedmann 1}) and (\ref{GBD Friedmann 2}) are just the Friedmann equations of the FRW universe in the generalized Brans-Dicke gravity by directly applying the FRW metric Eq.(\ref{FRW metric I}) to the gravitational field equation (\ref{Generalized Brans-Dicke 1}).
Specifically when $\omega(\phiup)\equiv\omega_{\text{BD}}$=constant and $V(\phiup)=0$ (and erase $G$ as $G\mapsto1/\phiup$ in standard Brans-Dicke), the thermodynamics-gravity correspondence just above reduces to the situation for the standard Brans-Dicke gravity \cite{Brans Dicke} and its FRW cosmology. Moreover, our results improves the setups of  $\{\rho_{\text{(MG)}}\,,P_{\text{(MG)}}\,,\bm\psi\,,W\cdots\}$ and the entropy production in
\cite{Cai II} and \cite{Cai III-2} for a similar  scalar-tensor theory with $\mathscr{L}=f(\phiup)R/(16\pi G)-\frac{1}{2}\nabla_{\alpha}\phiup \nabla^{\alpha}\phiup -V(\phiup)
+\mathscr{L}_m$.


\subsection{Equivalence between $f(R)$ and modified Brans-Dicke without kinetic term}

The two models analyzed just above have exhibited pretty similar behaviors. Next we consider a modified Brans-Dicke gravity $\mathscr{L}=\phiup\,R-V(\phiup)  +16\pi G\,\mathscr{L}_m$, which is just the Lagrangian density Eq.(\ref{Generalized Brans-Dicke with self-interaction potential})  in Sec.~\ref{Generalized Brans Dicke gravity with self-interaction potential} without the kinetic term $-\frac{\omega(\phiup)}{\phiup} \,\nabla_{\alpha}\phiup \nabla^{\alpha}$. Compare its field equation with that of the $\mathscr{L}=f(R)  +16\pi G\,\mathscr{L}_m$ gravity in
Sec.~\ref{fR gravity}:
\begin{equation}
\begin{split}
\phiup\,R_{\mu\nu}-\frac{1}{2}\Big(\phiup R-V(\phiup)\Big)\,g_{\mu\nu}+\,&\Big(g_{\mu\nu}\,\Box-\nabla_\mu\nabla_\nu\Big)\,\phiup\,=\,8\pi G\,T_{\mu\nu}^{(m)}\;,\\
f_R\,R_{\mu\nu}-\frac{1}{2} f(R)\, g_{\mu\nu}+\,&\Big(g_{\mu\nu}\,\Box-\nabla_\mu\nabla_\nu\Big) f_R\,=\,8\pi G\,T_{\mu\nu}^{(m)}\;.
\end{split}
\end{equation}
Clearly, these two field equations become identical with the following relations:
\begin{equation}\label{f(R) GBD equivalence}
f_R\,=\,\phiup \qquad\text{and}\qquad  f(R)=\phiup R-V(\phiup) \qquad\Rightarrow \qquad f_R\,R-f(R)\,=\,V(\phiup)\;.
\end{equation}
That is to say, the $f(R)$ fourth-order modified gravity in Sec.~\ref{fR gravity} and the generalized Brans-Dicke alternative gravity in Sec.~\ref{Generalized Brans Dicke gravity with self-interaction potential} are not totally independent. Instead, the former can be regarded as a subclass of the latter with vanishing coefficient $\omega(\phiup)\equiv0$ for the kinematic term $\nabla_{\alpha}\phiup \nabla^{\alpha}$, and the equivalence is built upon Eq.(\ref{f(R) GBD equivalence}). Applying the replacements  $f_R\mapsto \phiup$ and
$f_R\,R-f(R)\mapsto V(\phiup)$ to Sec.~\ref{fR gravity}, we obtain the modified-gravity SEM tensor as
\begin{equation}
T_{\mu\nu}^{\text{(MG)}}\,=\,\frac{1}{8\pi G}\,\Bigg(\big(\nabla_\mu\nabla_\nu-g_{\mu\nu}\Box\big)\,\phiup-\frac{1}{2}V\,g_{\mu\nu} \Bigg)\;,
\end{equation}
the energy density and pressure in $T^{\mu\,\text{(MG)}}_{\;\;\nu}$$=\text{diag}[-\rho_{\text{(MG)}},P_{\text{(MG)}},
P_{\text{(MG)}},P_{\text{(MG)}}]$ as
\begin{equation}
\rho_{\text{(MG)}}  \,=\,\frac{1}{8\pi G}\,\bigg(-3H\dot\phiup+\frac{1}{2}V\bigg)\quad\mbox{and}\quad
P_{\text{(MG)}}\,=\,\frac{1}{8\pi G}\,\bigg(\ddot{\phiup}+2H\dot \phiup-\frac{1}{2}V \bigg)\;,
\end{equation}
as well as the geometric mass $M$, nonequilibrium energy dissipation term $\mathcal{E}$,  horizon entropy $S$ and the nonequilibrium entropy production $d_pS$ to be
\begin{equation}
M\,=\,\frac{\phiup\Upsilon^3}{2G}\left( H^2+\frac{k}{a^2}\right)\quad,\quad
\mathcal {E}\,= \,\frac{4}{3}\pi\, \Upsilon^3\frac{\dot{\phiup}}{\phiup} \, \rho_{\text{eff}}\; dt  \quad,\quad
S\;\hat{=}\;\frac{A_{\text A}\phiup}{4G}\quad\text{and}\quad
d_p {S}\;\hat{=}\;-2\pi \Upsilon_{\text{A}}^2\frac{\dot{\phiup}}{G} \, dt\;.
\end{equation}
Finally the following equations are obtained from thermodynamics-gravity correspondence
\begin{equation}\label{GBD Friedmann 1 omega=0}
H^2+\frac{k}{a^2}\,=\,\frac{8\pi}{3}\frac{G}{\phiup}\,\rho_m + \frac{1}{3\phiup}\,\bigg(
-3H\dot\phiup+\frac{1}{2}V\bigg)\qquad\text{and}\qquad
\dot H-\frac{k}{a^2}\,=-4\pi\frac{G}{\phiup}\, \Big(\rho_m+P_m\Big)
-\frac{1}{2\phiup}\,\bigg(\ddot{\phiup}-H\dot \phiup\bigg)\;.
\end{equation}
It is easy to verify that, these thermodynamics quantities and equations precisely match  the generalized  Brans-Dicke in Sec.~\ref{Generalized Brans Dicke gravity with self-interaction potential} with $\omega(\phiup)\equiv0$.

Conversely, if start from these setups just above or those in Sec.~\ref{Generalized Brans Dicke gravity with self-interaction potential} with $\omega(\phiup)\equiv0$, the formulation in Sec.~\ref{fR gravity} can be recovered by applying the replacements $\phiup\mapsto f_R$ and
$V(\phiup)\mapsto f_R\,R-f(R)$.


\subsection{Scalar-tensor-chameleon gravity}\label{Scalar tensor chameleon gravity}

Consider the following Lagrangian density for the generic scalar-tensor-chameleon gravity \cite{scalar tensor chameleeon}  in the Jordan frame ,
\begin{equation}\label{Scalar-Tensor-Chameleon action}
\mathscr{L}_{\text{STC}}\,=\, F(\phiup)\,R-Z(\phiup) \nabla_{\alpha}\phiup \nabla^{\alpha}\phiup -2U(\phiup)
+16\pi G E(\phiup)\,\mathscr{L}_m \;,
\end{equation}
where $\{F(\phiup)\,, Z(\phiup)\,, E(\phiup)\}$ are arbitrary  functions of the scalar field $\phiup$, and $E(\phiup)$ is the chameleon function describing the coupling between $\phiup$ and the matter Lagrangian density $\mathscr{L}_m$. The name ``chameleon'' comes from the fact that in the presence of  $E(\phiup)$, the wave equation $\delta(\!\sqrt{-g}\,\mathscr{L}_{\text{STC}})/\delta \phiup=0$ of $\phiup$ becomes explicitly dependent on the matter content of the universe (eg. $\mathscr{L}_m$ or $T^{(m)}=g^{\mu\nu}T_{\mu\nu}^{(m)}$), which makes the wave equation change among different cosmic epoches as the dominant matter content varies \cite{scalar tensor chameleeon II}.
The gravitational field equation $\delta(\!\sqrt{-g}\,\mathscr{L}_{\text{STC}})/\delta g^{\mu\nu}=0$ is
\begin{equation}\label{STC eq 1}
R_{\mu\nu}-\frac{1}{2}R g_{\mu\nu} \,=\,8\pi G \frac{E(\phiup)}{F(\phiup)}\, T_{\mu\nu}^{(m)}
+\frac{1}{F(\phiup)}\Big(\nabla_\mu \nabla_\nu-g_{\mu\nu}\Box\Big)F(\phiup)
+\frac{Z(\phiup)}{F(\phiup)}\,\Big(\nabla_\mu\phiup \nabla_\nu \phiup-\frac{1}{2} g_{\mu\nu} \,\nabla_{\alpha}\phiup \nabla^{\alpha}\Big)
-\frac{U(\phiup)}{F(\phiup)} g_{\mu\nu}\;,
\end{equation}
so from the coefficient of  $T_{\mu\nu}^{(m)}$  we recognize
\begin{equation}
G_{\text{eff}}\,=\,\frac{E(\phiup)}{F(\phiup)}\,G\qquad\mbox{and}
\end{equation}
\begin{equation}
T_{\mu\nu}^{\text{(MG)}}\,=\,\frac{1}{8\pi G E(\phiup)}\,\Bigg(
\Big(\nabla_\mu \nabla_\nu-g_{\mu\nu}\Box\Big)F(\phiup)+Z(\phiup)\,\Big(\nabla_\mu\phiup \nabla_\nu \phiup-\frac{1}{2} g_{\mu\nu} \,\nabla_{\alpha}\phiup \nabla^{\alpha}\phiup\Big)
-U(\phiup) g_{\mu\nu} \Bigg)\;.
\end{equation}
Note that  \cite{scalar tensor chameleeon} however adopted $G_{\text{eff}}=G/F(\phiup)$ to study the second law of thermodynamics for the flat FRW universe, the chameleon function $E(\phiup)$ excluded from $G_{\text{eff}}$. Substituting the FRW metric Eq.(\ref{Friedmann eqn 1st}) into $T_{\mu\nu}^{\text{(MG)}}$, the energy density and pressure for  $T^{\mu\,\text{(MG)}}_{\;\;\nu}=\text{diag}[-\rho_{\text{(MG)}},P_{\text{(MG)}},$ $
P_{\text{(MG)}},P_{\text{(MG)}}]$  are found to be
\begin{equation}\label{STC rho P}
\begin{split}
\hspace{-3mm}\rho_{\text{(MG)}}\,=\, \frac{1}{ 8\pi G\,E(\phiup)}\bigg(-3H\dot F+\frac{1}{2}\,Z(\phiup)\,\dot{\phiup}^2+U\bigg)
\quad\text{and}\quad
P_{\text{(MG)}}\,=\,\frac{1}{ 8\pi G\,E(\phiup)}\bigg(\ddot F+2H\dot F +\frac{1}{2} Z(\phiup)\,\dot{\phiup}^2 -U\bigg) \;,
\end{split}
\end{equation}
where the compact notations $\dot F$ and $\ddot F$ can be replaced by $F_\phiup \,\dot\phiup$ and $F_\phiup\,\ddot \phiup+F_{\phiup\phiup}\, \dot{\phiup}^2$, respectively. As $G_{\text{eff}}=G E(\phiup)/F(\phiup)$, the Hawking or Misner-Sharp geometric mass becomes
\begin{equation}\label{STC mass}
M\,=\,\frac{F(\phiup)\Upsilon^3}{2GE(\phiup)}\left( H^2+\frac{k}{a^2}\right)\qquad\text{with}\qquad
M_{\text A}\;\hat{=}\;\frac{F(\phiup)\Upsilon_{\text A}}{2GE(\phiup)}\;,
\end{equation}
while the nonequilibrium energy dissipation $\mathcal {E}$ in the conservation equation and the Wald-Kodama entropy of the horizon $S$ are respectively
\begin{equation}\label{STC dissipation}
\mathcal{E}\,= \,\frac{4}{3}\pi\, \Upsilon^3 \frac{G}{F(\phiup)^2}\Big(E(\phiup)\dot F -F(\phiup)\dot E \Big) \, \rho_{\text{eff}}\; dt
\qquad \text{and} \qquad S\,=\,\frac{A_{\text A}F(\phiup)}{4GE(\phiup)}\;,
\end{equation}
where in $\mathcal{E}$ the compact notation $E(\phiup)\dot F -F(\phiup)\dot E$ can be expanded into $(EF_\phiup-FE_\phiup)\,\dot\phiup$. Moreover, using the unified formulation developed in Sec.~\ref{Inside the cosmological apparent horizon} and Sec.~\ref{On the cosmological apparent horizon}, for the interior and the horizon  we obtain
\begin{equation}
H^2+\frac{k}{a^2}\,=\,
\frac{8\pi}{3} \frac{G E(\phiup)}{F(\phiup)}\, \rho_m
+\frac{1}{3F(\phiup)}\bigg(-3H\dot F+\frac{1}{2}\,Z(\phiup)\,\dot{\phiup}^2+U\bigg) \;,
\end{equation}
\begin{equation}
\hspace{0mm}\dot H-\frac{k}{a^2}\,=-4\pi\frac{G E(\phiup)}{F(\phiup)}\, \Big(\rho_m+P_m\Big)
-\frac{1}{2F(\phiup)}\,\bigg(\ddot F-H\dot F +Z(\phiup)\,\dot{\phiup}^2 \bigg)\;.
\end{equation}
With $\dot F=F_\phiup \dot\phiup$ and $\ddot F=F_\phiup\ddot \phiup+F_{\phiup\phiup}\, \dot{\phiup}^2$,
they can be recast into
\begin{equation}\label{STC Friedmann 1}
H^2+\frac{k}{a^2}\,=\,
\frac{8\pi}{3} \frac{G E(\phiup)}{F(\phiup)}\, \rho_m
+\frac{1}{3F(\phiup)}\bigg(-3HF_\phiup\,\dot\phiup+\frac{1}{2}Z(\phiup)\,\dot{\phiup}^2+U\bigg) \;,
\end{equation}
\begin{equation}\label{STC Friedmann 2}
\hspace{0mm}\dot H-\frac{k}{a^2}\,=-4\pi\frac{G E(\phiup)}{F(\phiup)}\, \Big(\rho_m+P_m\Big)
-\frac{1}{2F(\phiup)}\,\bigg(F_\phiup\,\ddot\phiup+F_{\phiup\phiup}\,\dot{\phiup}^2-HF_\phiup\,\dot\phiup
+Z(\phiup)\,\dot{\phiup}^2 \bigg)\;.
\end{equation}
At the same time, the nonequilibrium entropy production turns out to be
\begin{equation}\label{STC entropy production}
d_p {S}\,=\,2\pi \Upsilon_{\text{A}}^2\, \frac{1}{GE(\phiup)^2}\Big(FE_\phiup-EF_\phiup\Big)\,\dot\phiup \, dt\;.
\end{equation}
We have verified by direct substitution of the FRW metric Eq.(\ref{FRW metric I})
into Eq.(\ref{STC eq 1}) that Eqs.(\ref{STC Friedmann 1}) and (\ref{STC Friedmann 1}) are indeed the Friedmann equations of the FRW universe in the
scalar-tensor-chameleon gravity.

Compare the scalar-tensor-chameleon theory with the generalized Brans-Dicke gravity in Sec.~\ref{Generalized Brans Dicke gravity with self-interaction potential}, and we find that besides the nonminimal coupling $F(\phiup)R$ in the Lagrangian density, the chameleon field $E(\phiup)$ coupled to $\mathscr{L}_m$ causes extra nonequilibrium energy dissipation and entropy production, as shown by Eqs.(\ref{STC dissipation}) and (\ref{STC entropy production}). On the other hand, in the absence of the chameleon function, $E(\phiup)\equiv 1$, $E_\phiup=0$, and with $F(\phiup)\mapsto \phiup$, $F_\phiup\mapsto 1$, $F_{\phiup\phiup}\mapsto 0$, $Z(\phiup)\mapsto \omega(\phiup)/\phiup$, $U\mapsto \frac{1}{2}V$, we recover the generalized  Brans-Dicke in Sec.~\ref{Generalized Brans Dicke gravity with self-interaction potential}.

In \cite{Misner-Sharp mass III}, for the scalar-tensor gravity $\mathscr{L}=F(\phiup)R/(16\pi G)-\frac{1}{2}\nabla_{\alpha}\phiup \nabla^{\alpha}\phiup -V(\phiup) +\mathscr{L}_m$\,, the generalized Misner-Sharp mass/energy in the FRW universe is found to be
\begin{equation}\label{Cai MS mass scalar tensor}
E_{\text{eff}}\,=\,\frac{\Upsilon^3}{2G}\,\Bigg( F(\phiup)\,\Big( H^2+\frac{k}{a^2}\Big)+
H\dot F-\frac{4\pi}{3}\,\Big( \frac{1}{2}\,\dot{\phiup}^2+V\Big)\,\Bigg)\;.
\end{equation}
(Note: A typo in Eq.(A8) of \cite{Misner-Sharp mass III} is corrected here by  either checking the derivation of Eq.(A8), or by referring to  Eq.(\ref{Cai MS mass fR}) with the correspondence $f_R=\phiup$ and $f_R\,R-f(R)=V$ as in Eq.(\ref{f(R) GBD equivalence}), despite the nonzero kinetic term $-\frac{1}{2}\nabla_{\alpha}\phiup \nabla^{\alpha}\phiup$.) Compared with Eq.(\ref{Scalar-Tensor-Chameleon action}),
\cite{Misner-Sharp mass III} actually adopts a different scaling convention for the Lagrangian density; in accordance with Eq.(\ref{Scalar-Tensor-Chameleon action}), we rescale \cite{Misner-Sharp mass III} by
\begin{equation}\label{Cai MS mass L}
\mathscr{L}=F(\phiup)R-\frac{1}{2}\nabla_{\alpha}\phiup \nabla^{\alpha}\phiup -V(\phiup) +16\pi G\mathscr{L}_m\;,
\end{equation}
and consequently
\begin{equation}\label{Cai MS mass scalar tensor II}
E_{\text{eff}}\,=\,\frac{\Upsilon^3}{2G}\,\Bigg( F(\phiup)\,\Big( H^2+\frac{k}{a^2}\Big)+
H\dot F-\frac{1}{6}\,\Big( \frac{1}{2}\,\dot{\phiup}^2+V\Big)\,\Bigg)\;,
\end{equation}
which can be expanded into
\begin{equation}\label{Cai MS mass scalar tensor III}
\begin{split}
E_{\text{eff}}\,=\,\frac{F(\phiup)\Upsilon^3}{2G}\,\Big( H^2+\frac{k}{a^2}\Big)-
\frac{4}{3}\pi\Upsilon^3\cdot\frac{1}{8\pi G}\, \Bigg(
-3H\dot F+  \frac{1}{4}\,\dot{\phiup}^2+\frac{1}{2}\,V \,\Bigg)\;.
\end{split}
\end{equation}
As a subclass of the generic scalar-tensor-chameleon gravity Eq.(\ref{Scalar-Tensor-Chameleon action}) with $E(\phiup)\mapsto 1$\,, $Z(\phiup)\mapsto \frac{1}{2}$ and $U\mapsto \frac{1}{2}V$ for the Lagrangian density Eq.(\ref{Cai MS mass L}), the energy density $\rho_{\text{(MG)}}$ in Eq.(\ref{STC rho P}) and the mass $M$ in Eq.(\ref{STC mass}) reduce to become
\begin{equation}
\begin{split}
\rho_{\text{(MG)}}\,=\, \frac{1}{ 8\pi G}\bigg(-3H\dot F+\frac{1}{4}\,\dot{\phiup}^2+\frac{1}{2}V\bigg)
\quad\text{and}\quad
M\,=\,\frac{F(\phiup)\Upsilon^3}{2G}\left( H^2+\frac{k}{a^2}\right) \;,
\end{split}
\end{equation}
which finally recast Eq.(\ref{Cai MS mass scalar tensor III}) into
\begin{equation}\label{Cai MS mass scalar tensor III}
\begin{split}
E_{\text{eff}}\,=\,M-\rho_{\text{(MG)}}V\,=\,\Big(\rho_{m}+\rho_{\text{(MG)}}\Big)\,V-\rho_{\text{(MG)}}V
\,=\,\rho_{m} V\;.
\end{split}
\end{equation}
Hence, the ``generalized Misner-Sharp energy $E_{\text{eff}}$'' for the FRW universe within the scalar-tensor gravity in \cite{Misner-Sharp mass III} is
in fact the pure Misner-Sharp mass of physical matter for the same gravity in our work, just like the case of $f(R)$ gravity in Sec.~\ref{fR gravity}.


\subsection{Reconstruction of the physical mass $\rho_m V$ in generic modified gravity}

Before proceeding to analyze more examples, we would like to give some remarks on the problem of reconstructing physical mass.
Recall that in GR  the mass $\rho_m V$ of the physical matter (like baryon dust, radiation) can be geometrically recovered by the Hawking mass for twist-free spacetimes \cite{Hawking mass} and the Misner-Sharp mass for spherically symmetric spacetimes \cite{Misner-Sharp mass}.
In modified gravity, the physical matter content determines the FRW spacetime geometry Eq.(\ref{FRW metric I}) through more generic field equations which usually contain nonlinear and higher-order curvature terms beyond GR. Thus, how to reconstruct the mass of the physical matter from the spacetime geometry?

In \cite{Misner-Sharp mass III}, Cai el al. generalized the Misner-Sharp mass of GR into higher-dimensional Gauss-Bonnet gravity and the $f(R)$ (plus the scalar-tensor FRW) gravity in four dimensions. As just shown in Sec.\ref{fR gravity} and Sec.~\ref{Scalar tensor chameleon gravity}, for the FRW universe the results in \cite{Misner-Sharp mass III} do match the physical material mass $\rho_m V$ in our unified formulation. In fact, for the FRW universe governed by generic modified gravity with the field equation $R_{\mu\nu}-Rg_{\mu\nu}/2=8\pi G_{\text{eff}}  T_{\mu\nu}^{\text{(eff)}}$, the mass $\mathcal{M}^{(m)}=\rho_m V$ of the physical matter content can be reconstructed from an geometric approach by
\begin{equation}\label{mass reconstruction}
\mathcal{M}^{(m)}\,=\,\frac{\Upsilon^3}{2G_{\text{eff}}}\left( H^2+\frac{k}{a^2}\right)-\frac{4\pi \Upsilon^3}{3}\rho_{\text{(MG)}}\,,
\end{equation}
where $\rho_{\text{(MG)}}$ is the density of modified-gravity effects collecting the nonlinear and higher-order geometric terms and joining $T^{\mu\,\text{(MG)}}_{\;\;\nu}=\text{diag}[-\rho_{\text{(MG)}},P_{\text{(MG)}},
P_{\text{(MG)}},P_{\text{(MG)}}]$, as concretely shown just before for $f(R)$, generalized Brans-Dicke and scalar-tensor-chameleon gravity.
When going beyond the FRW geometry in modified gravity, however, the validity of
\begin{equation}\label{Hawking mass reconstruction}
\begin{split}
\mathcal{M}_{\text{Hk}}^{(m)}\; &=\frac{1}{4\pi G_{\text{eff}}} \left( \int  \frac{dA}{4\pi} \right)^{\frac{1}{2}}\int \Big( -\Psi_2-\sigma_{\text{NP}} \lambda_{\text{NP}} +\Phi_{11}+\Lambda_{\text{NP}} \Big)\,dA -\frac{4\pi \Upsilon^3}{3}\rho_{\text{(MG)}}\\
&= \frac{1}{4\pi G_{\text{eff}}} \left( \int  \frac{dA}{4\pi} \right)^{\frac{1}{2}} \Bigg( 2\pi-\int  \rho_{\text{NP}}\,\mu_{\text{NP}}\,dA  \Bigg)-\frac{4\pi \Upsilon^3}{3}\rho_{\text{(MG)}}
\end{split}
\end{equation}
to recover the physical mass  $\rho_m V$  for an arbitrary twist-free spacetime based on the effective Hawking mass Eq.(\ref{Hawking mass}) in our unified formulation, and the feasibility of
\begin{equation}\label{Misner-Sharp mass reconstruction}
 \mathcal{M}_{\text{MS}}^{(m)}\;=\,\frac{\Upsilon }{2G_{\text{eff}}} \,
 \Big( 1- h^{\alpha\beta}\partial_\alpha \Upsilon\partial_\beta \Upsilon\Big)-\frac{4\pi \Upsilon^3}{3}\rho_{\text{(MG)}}\,,
\end{equation}
for generic spherically symmetric spacetimes based on the effective Misner-Sharp mass Eq.(\ref{Misner-Sharp mass}),  remain to be examined.


\subsection{Quadratic gravity}\label{Quadratic gravity}

For quadratic gravity \cite{Example Quardratic gravity second paper}, the  Lagrangian density  is constructed by combining the Hilbert-Einstein density of GR with the linear superposition of some well-known quadratic (as opposed to cubic and quartic) algebraic curvature invariants such as $R^2$, $R_{\mu\nu}R^{\mu\nu}$, $S_{\mu\nu}S^{\mu\nu}$ (with $S_{\mu\nu}\coloneqq R_{\mu\nu}-\frac{1}{4}\,R\,g_{\mu\nu}$), $R_{\mu\alpha\nu\beta}R^{\mu\alpha\nu\beta}$, $C_{\mu\alpha\nu\beta}C^{\mu\alpha\nu\beta}$ (Weyl tensor square), say $\mathscr{L}=R+a\,R^2+b\, R_{\mu\nu}R^{\mu\nu}+c\, S_{\mu\nu}S^{\mu\nu}+ d\, R_{\mu\alpha\nu\beta}R^{\mu\alpha\nu\beta}+ e\, C_{\mu\alpha\nu\beta}C^{\mu\alpha\nu\beta} +16\pi G \mathscr{L}_m$ where $\{a,b,c,d,e\}$ are real-valued constants. However, these quadratic invariants are not totally independent of each other, as $S_{\mu\nu}S^{\mu\nu} = R_{\mu\nu}R^{\mu\nu}-\frac{1}{4}\,R^2 $, $C_{\mu\alpha\nu\beta}C^{\mu\alpha\nu\beta}=R_{\mu\alpha\nu\beta}R^{\mu\alpha\nu\beta}-2R_{\mu\nu}R^{\mu\nu}+R^2/3$, and moreover $R_{\mu\alpha\nu\beta}R^{\mu\alpha\nu\beta}$ can be absorbed into the Gauss-Bonnet invariant
$\mathcal{G}\,\coloneqq \, R^2-4R_{\mu\nu}R^{\mu\nu}+R_{\mu\alpha\nu\beta}R^{\mu\alpha\nu\beta}$ which does not contribute to the field equation since $\delta\int d^4x\sqrt{-g}\,\mathcal{G}/\delta g^{\mu\nu}\equiv 0$ (eg. \cite{AA Tian-Booth Paper}). Hence, it is sufficient to consider the following Lagrangian density for quadratic gravity
\begin{equation}
\mathscr{L}_{\text{QG}}\,=\,  R+ a\, R^2 +b \, R_{\mu\nu}R^{\mu\nu}  + 16\pi G \mathscr{L}_m \;,
\end{equation}
and the field equation is \cite{AA Tian-Booth Paper}
\begin{equation}\label{Quadratic FieldEqn QRc}
\begin{split}
-\frac{1}{2}\big(  R+ a\cdot R^2 +b\cdot R_c^2 \big)\,g_{\mu\nu}
+\big(1+2aR\big)\,R_{\mu\nu}
+2a\,\big(g_{\mu\nu}\Box-\nabla_\mu\!\nabla_\nu\big)\,R+b\cdot H_{\mu\nu}^{\text{(QG)}} \,=\,8\pi G\,
T_{\mu\nu}^{(m)}\;,
\end{split}
\end{equation}
where $R_c^2$ is the straightforward abbreviation for the Ricci tensor square $R_{\mu\nu}R^{\mu\nu}$ to shorten some upcoming expressions below, and
\begin{equation}
H_{\mu\nu}^{\text{(QG)}}\,=\,2 R_{\mu\alpha\nu\beta}R^{\alpha\beta}
+ \big(\frac{1}{2}\,g_{\mu\nu}\Box -\nabla_\mu \nabla_\nu \big)\,R+ \Box R_{\mu\nu}\;.
\end{equation}
It can be rewritten into
\begin{equation}\label{Quadratic FieldEqn QRc II}
\begin{split}
R_{\mu\nu}-\frac{1}{2}Rg_{\mu\nu}
\,=\,8\pi \frac{G}{1+2aR}\,\Big(T_{\mu\nu}^{(m)}
+T_{\mu\nu}^{\text{(MG)}}\Big)
\end{split}
\end{equation}
where
\begin{equation}
G_{\text{eff}}\,=\,\frac{G}{1+2aR}\qquad\text{and}\quad
\end{equation}
\begin{equation}\label{Quadratic FieldEqn QRc III}
\begin{split}
T_{\mu\nu}^{\text{(MG)}}\,=\,
\frac{1}{8\pi G}\,\Bigg(\frac{1}{2}\Big( b\cdot R_c^2  -aR^2 \Big)\,g_{\mu\nu}+(2a+b)\,\nabla_\mu\!\nabla_\nu R-\big(2a+\frac{b}{2}\big)\,g_{\mu\nu}\Box\, R
-2b\,\Big(2 R_{\mu\alpha\nu\beta}R^{\alpha\beta} +\Box R_{\mu\nu}\Big) \Bigg)\;.
\end{split}
\end{equation}
Substitute the FRW metric Eq.(\ref{FRW metric I}) into $T_{\mu\nu}^{\text{(MG)}}$, and with $T^{\mu\,\text{(MG)}}_{\;\;\nu}=\text{diag}[-\rho_{\text{(MG)}},P_{\text{(MG)}},$ $P_{\text{(MG)}},P_{\text{(MG)}}]$ we get
\begin{equation}
\rho_{\text{(MG)}}\,=\,\frac{1}{8\pi G}\,\Bigg(\frac{a}{2}R^2- \frac{b}{2}R_c^2 +\frac{b}{2}\,\ddot R-\big(4a+b\big)\, H\dot R
+4b\, R^t_{\;\;\alpha t\beta} +2b\,\Box R_t^{\;\;t} \Bigg)\;,
\end{equation}
\begin{equation}
\begin{split}
P_{\text{(MG)}}\,=\,
\frac{1}{8\pi G}\,\Bigg( \frac{b}{2}R_c^2  -\frac{a}{2}R^2 +\big(2a+\frac{b}{2}\big)\,\ddot{R}+ \big(4a+\frac{b}{2}\big)\, H\dot{R}
-4b\,R^r_{\;\;\alpha r\beta}R^{\alpha\beta} -2b\,\Box R_r^{\;\;r} \Bigg)\;.
\end{split}
\end{equation}
where we have used $R^t_{\;\;\alpha t\beta}=-R_{t\alpha t\beta}$ and $\Box R_t^{\;\;t}=-\Box R_{tt}$ in $\rho_{\text{(MG)}}$ under the FRW metric Eq.(\ref{FRW metric I}).
Also, since $G_{\text{eff}}=G/\phiup$, the geometric mass enclosed in a sphere of radius $\Upsilon$ is
\begin{equation}
M\,=\,\frac{(1+2aR)\,\Upsilon^3}{2G}\left( H^2+\frac{k}{a^2}\right)\qquad\text{with}\qquad
M_{\text A}\;\hat{=}\;\frac{(1+2aR)\,\Upsilon_{\text A}}{2G}\;,
\end{equation}
while the nonequilibrium energy dissipation $\mathcal {E}$ associated with the evolution of $G_{\text{eff}}$ and the Wald-Kodama entropy $E$ of the horizon are respectively
\begin{equation}
\mathcal {E}\,= \,\frac{4}{3}\pi\, \Upsilon^3 \frac{2a \dot{R}}{1+2aR} \, \rho_{\text{eff}}\; dt  \qquad\text{and}\qquad
S\,=\,\frac{A_{\text A}\,(1+2aR)}{4G}\;.
\end{equation}
Following the unified formulation developed in Sec.~\ref{Inside the cosmological apparent horizon} and Sec.~\ref{On the cosmological apparent horizon} to study $dM=dE=A\bm\psi+WdV+\mathcal{E}$ for the region $\Upsilon\leq \Upsilon_{\text A}$ and $T(dS+d_PS)\,\hat{=}\,\delta Q\,\hat{=}-(A_{\text A}\bm\psi+\mathcal{E}_{\text A})$ for the horizon itself, we find
\begin{equation}\label{QG Friedmann 1}
H^2+\frac{k}{a^2}\,=\,\frac{8\pi}{3}\frac{G}{1+2aR}\,\rho_m + \frac{1}{3(1+2aR)}\,\Bigg(\frac{a}{2}R^2- \frac{b}{2}R_c^2 +\frac{b}{2}\,\ddot R-\big(4a+b\big)\, H\dot R
+4b\, R^t_{\;\;\alpha t\beta} +2b\,\Box R_t^{\;\;\,t} \Bigg)
\end{equation}
\begin{equation}\label{QG Friedmann 2}
\hspace{0mm}\dot H-\frac{k}{a^2}\,=-4\pi\frac{G}{1+2aR} \Big(\rho_m+P_m\Big)
-\frac{1}{2(1+2aR)}\,\Bigg( \big(2a+b\big)\,\ddot{R}-\frac{b}{2}\, H\dot{R}
+4b\,(R^t_{\;\;\alpha t\beta}-R^r_{\;\;\alpha r\beta})R^{\alpha\beta} +2b\,\Box \big(R_t^{\;\;\,t}-R_r^{\;\;r}\big) \Bigg)\;,
\end{equation}
while the  nonequilibrium entropy production on the horizon is
\begin{equation}
d_p {S}\;\hat{=}\;
-4\pi \Upsilon_{\text{A}}^2\,\frac{a \dot{R}}{G} \, dt\;.
\end{equation}
We have verified that the thermodynamic relations Eqs.(\ref{QG Friedmann 1}) and (\ref{QG Friedmann 2})
are equivalent to the gravitational Friedmann equations by substituting the FRW metric  Eq.(\ref{FRW metric I})
into the quadratic field equations (\ref{Quadratic FieldEqn QRc II})
and (\ref{Quadratic FieldEqn QRc III}).

Just like the treatment of $f(R)$ gravity in Sec.~\ref{fR gravity},  to keep the expressions of
$\rho_{\text{(MG)}}$\,, $P_{\text{(MG)}}$ and the Friedmann equations (\ref{QG Friedmann 1}) and (\ref{QG Friedmann 2}) clear and readable,
we continue using compact notations for $R$\,, $R_c^2$\,, $\dot R$\,, $\ddot R$\,, $R^t_{\;\;\alpha t\beta}R^{\alpha\beta}$\,,  $R^r_{\;\;\alpha r\beta}R^{\alpha\beta}$\,, $\Box R_t^{\;\;\,t}$ and $\Box R_r^{\;\;\,r}$\,, and one should keep in mind that for the FRW metric Eq.(\ref{FRW metric I}), these quantities are already known and can be fully expanded into higher-derivative and nonlinear terms of $H$ or $a$.


\subsection{$f(R,\mathcal{G})$ generalized Gauss-Bonnet gravity}\label{fRG gravity}

The generalized Gauss-Bonnet gravity under discussion is given by the Lagrangian density $\mathscr{L}_{\text{GB}}=f(R,\mathcal{G})+16\pi G\mathscr{L}_m$ \cite{Example GaussBonnet second model f(R G)+Lm} where $\mathcal{G}=R^2-4R_{\mu\nu}R^{\mu\nu}+R_{\mu\alpha\nu\beta}R^{\mu\alpha\nu\beta}$ is the Gauss-Bonnet invariant. This is in fact a subclass of the $\mathscr{L}=f(R\,,R_{\mu\nu}R^{\mu\nu}\,,R_{\mu\alpha\nu\beta}R^{\mu\alpha\nu\beta})+16\pi G\mathscr{L}_m$ gravity \cite{Example Carroll R+ f(R Rc2 Rm2)+2kLm} with explicit dependence on $R^2$ and satisfying the ``coherence condition'' $f_{R^2}=f_{R_m^2}= -f_{R_c^2}/4$ \cite{AA Tian-Booth Paper} ($R_m^2$ and $R_c^2$ are the intuitive abbreviations for the Riemann tensor square $R_{\mu\alpha\nu\beta}R^{\mu\alpha\nu\beta}$ and the Ricci tensor square $R_{\mu\nu}R^{\mu\nu}$, respectively).
The field equation for $f(R,\mathcal{G})$ gravity reads
\begin{equation}\label{GB Field Eqn}
\begin{split}
R_{\mu\nu}-\frac{1}{2}R g_{\mu\nu}\;=\;8\pi \frac{G}{f_R+2Rf_{\mathcal{G}}}\, T_{\mu\nu}^{(m)}
+\big(f_R+2Rf_{\mathcal{G}}\big)^{-1}\,\Bigg(\frac{1}{2}\,\Big(f-(f_R+2Rf_{\mathcal{G}})\,R\,\Big)\,g_{\mu\nu}&\\
+\big(\nabla_\mu\!\nabla_\nu-g_{\mu\nu}\Box\big)\,f_R
+2R\,\big(\nabla_\mu\!\nabla_\nu-g_{\mu\nu}\Box\big)\,f_{\mathcal{G}}
+4R_{\mu\nu}\Box f_{\mathcal{G}}+H_{\mu\nu}^{\text{(GB)}}&\Bigg)\;,
\end{split}
\end{equation}
where
\begin{equation}\label{GB Field Eqn II}
\begin{split}
H_{\mu\nu}^{\text{(GB)}} \,\coloneqq\;
& 4f_{\mathcal{G}}\!\cdot\! R_\mu^{\;\;\,\alpha}R_{\alpha\nu}+4f_{\mathcal{G}}\!\cdot\!R_{\mu\alpha\nu\beta}R^{\alpha\beta} -2f_{\mathcal{G}}\!\cdot\! R_{\mu\alpha\beta\gamma}R_{\nu}^{\;\;\,\alpha\beta\gamma}-4R_{\mu}^{\;\;\,\alpha}\nabla_\alpha\!\nabla_{\nu}f_{\mathcal{G}} \\
-&4R_{\nu}^{\;\;\,\alpha}\nabla_\alpha\!\nabla_{\mu} f_{\mathcal{G}}
+4g_{\mu\nu} \!\cdot\! R^{\alpha\beta}\nabla_\alpha\!\nabla_\beta f_{\mathcal{G}}-4\,R_{\alpha\mu \beta\nu} \nabla^\beta \nabla^\alpha  f_{\mathcal{G}}\;,
\end{split}
\end{equation}
and $\{f,f_R,f_{\mathcal{G}}=\partial f/\partial\mathcal{G}\}$ are all functions of $(R,\mathcal{G})$. Note that in $H_{\mu\nu}^{\text{(GB)}}$ the second-order-derivative operators $\{\Box,\nabla_\alpha\!\nabla_\nu, \text{etc}\}$  only act on the scalar functions $f_{\mathcal{G}}$.
Hence,
\begin{equation}\label{GB Geff}
G_{\text{eff}}\,=\,\frac{G}{f_R+2Rf_{\mathcal{G}}}\qquad\text{and}\quad
\end{equation}
\begin{equation}\label{GB TMG}
\begin{split}
T_{\mu\nu}^{\text{(MG)}}\,=\,
\frac{1}{8\pi G}\,\Bigg(\frac{1}{2}\,\Big(f-(f_R+2Rf_{\mathcal{G}})\,R\,\Big)\,g_{\mu\nu}
+\big(\nabla_\mu\!\nabla_\nu-g_{\mu\nu}\Box\big)\,f_R
+2R\,\big(\nabla_\mu\!\nabla_\nu-g_{\mu\nu}\Box\big)\,f_{\mathcal{G}}
+4R_{\mu\nu}\Box f_{\mathcal{G}}+H_{\mu\nu}^{\text{(GB)}}\Bigg)\;.
\end{split}
\end{equation}
Substitute the FRW metric Eq.(\ref{FRW metric I}) into $T_{\mu\nu}^{\text{(MG)}}$ with $T^{\mu\,\text{(MG)}}_{\;\;\nu}=\text{diag}[-\rho_{\text{(MG)}},P_{\text{(MG)}},$ $P_{\text{(MG)}},P_{\text{(MG)}}]$, and in  compact notations we obtain
\begin{equation}
\rho_{\text{(MG)}}\,=\,\frac{1}{8\pi G}\,\Bigg( \frac{1}{2}\,(f_R+2Rf_{\mathcal{G}})\,R-\frac{1}{2}\,f-3H\dot{f}_R-6RH\dot{f}_{\mathcal{G}}
+4R_t^{\;\;t}(\ddot{f}_{\mathcal{G}}+3H\dot{f}_{\mathcal{G}})-H_{t\,\text{(GB)}}^{\;\;t}
\Bigg)\;,
\end{equation}
\begin{equation}
\begin{split}
P_{\text{(MG)}}\,=\,
\frac{1}{8\pi G}\,\Bigg(\frac{1}{2}\,f- \frac{1}{2}\,(f_R+2Rf_{\mathcal{G}})\,R
+\ddot{f}_R+2H\dot{f}_R+2R(\ddot{f}_{\mathcal{G}}+2H\dot{f}_{\mathcal{G}})
-4R_r^{\;\;r}(\ddot{f}_{\mathcal{G}}+3H\dot{f}_{\mathcal{G}})+H_{r\,\text{(GB)}}^{\;\;r}
\Bigg)\;,
\end{split}
\end{equation}
where we have used the properties $R_t^{\;\;t}=-R_{tt}$ and $H_{t\,\text{(GB)}}^{\;\;t}=-H_{tt}^{\text{(GB)}}$
in $\rho_{\text{(MG)}}$ under the FRW metric Eq.(\ref{FRW metric I}).
Since $G_{\text{eff}}=G/(f_R+2Rf_{\mathcal{G}})$, the geometric mass within a sphere of radius $\Upsilon$ is
\begin{equation}
M\,=\,\frac{(f_R+2Rf_{\mathcal{G}})\,\Upsilon^3}{2G}\left( H^2+\frac{k}{a^2}\right)\qquad\text{with}\qquad
M_{\text A}\;\hat{=}\;\frac{(f_R+2Rf_{\mathcal{G}})\,\Upsilon_{\text A}}{2G}\;,
\end{equation}
while the nonequilibrium energy dissipation $\mathcal {E}$ associated with the evolution of $G_{\text{eff}}$ and the Wald-Kodama entropy $S$ of the horizon are respectively
\begin{equation}
\mathcal {E}\,= \,\frac{4}{3}\pi\, \Upsilon^3 \frac{\dot{f}_R+2\dot{R}f_{\mathcal{G}}+2R\dot{f}_{\mathcal{G}}}{f_R+2Rf_{\mathcal{G}}} \, \rho_{\text{eff}}\; dt  \qquad\text{and}\qquad
S\,=\,\frac{A_{\text A}\,(f_R+2Rf_{\mathcal{G}})}{4G}\;.
\end{equation}
Following the unified formulation developed in Sec.~\ref{Inside the cosmological apparent horizon} and Sec.~\ref{On the cosmological apparent horizon} to study $dM=dE=A\bm\psi+WdV+\mathcal{E}$ for the region $\Upsilon\leq \Upsilon_{\text A}$ and $T(dS+d_PS)\,\hat{=}\,\delta Q\,\hat{=}-(A_{\text A}\bm\psi+\mathcal{E}_{\text A})$ for the horizon itself, we find
\begin{equation}\label{GB Friedmann 1}
\begin{split}
H^2+\frac{k}{a^2}\,=\,\frac{8\pi}{3}\frac{G}{f_R+2Rf_{\mathcal{G}}}\,\rho_m + \frac{1}{3\,\big(f_R+2Rf_{\mathcal{G}}\big)}\,\Bigg( \frac{1}{2}\,\big(f_R+2Rf_{\mathcal{G}}\big)\,R-\frac{1}{2}f-3H\,\big(\dot{f}_R+2R\dot{f}_{\mathcal{G}}\big)&\\
+4R_t^{\;\;t}\,\big(\ddot{f}_{\mathcal{G}}+3H\dot{f}_{\mathcal{G}}\big)-H_{t\,\text{(GB)}}^{\;\;t}&
\Bigg)\;,
\end{split}
\end{equation}
\begin{equation}\label{GB Friedmann 2}
\begin{split}
\hspace{0mm}\dot H-\frac{k}{a^2}\,=-4\pi\frac{G}{f_R+2Rf_{\mathcal{G}}} \Big(\rho_m+P_m\Big)
-\frac{1}{2\,\big(f_R+2Rf_{\mathcal{G}}\big)}\,\Bigg( \ddot{f}_R-H\dot{f}_R
+2R\ddot{f}_{\mathcal{G}}-2RH\dot{f}_{\mathcal{G}}&\\
+4\,\big(R_t^{\;\;t}-R_r^{\;\;r}\big)\,\big(\ddot{f}_{\mathcal{G}}+3H\dot{f}_{\mathcal{G}}\big)
-H_{t\,\text{(GB)}}^{\;\;t}+H_{r\,\text{(GB)}}^{\;\;r}&
\Bigg)\;,
\end{split}
\end{equation}
while the  nonequilibrium entropy production on the horizon is
\begin{equation}
d_p {S}\;\hat{=}\;
-2\pi \Upsilon_{\text{A}}^2\,\frac{\dot{f}_R+2\dot{R}f_{\mathcal{G}}+2R\dot{f}_{\mathcal{G}}}{G} \, dt\;.
\end{equation}
We have verified that the thermodynamic relations Eqs.(\ref{GB Friedmann 1}) and (\ref{GB Friedmann 2})
are really the gravitational Friedmann equations by substituting the FRW metric  Eq.(\ref{FRW metric I})
into the generalized Gauss-Bonnet field equations (\ref{GB Field Eqn})
and (\ref{GB Field Eqn II}). Moreover, by setting $f_{\mathcal{G}}=0$ and thus $\dot{f}_{\mathcal{G}}=\ddot{f}_{\mathcal{G}}=0$\,, the situation of the $f(R\,,{\mathcal{G}})$ generalized Gauss-Bonnet gravity reduces to become the case of $f(R)$ gravity in Sec.~\ref{fR gravity}.


\subsection{Self-inconsistency of $f(R,\mathcal{G})$ gravity}\label{Self-consistency of f(RG) gravity}

The $f(R,\mathcal{G})$ example just above is based on Eqs.(\ref{GB Field Eqn}) and (\ref{GB Field Eqn II}), which together with their contravariant forms constitute the standard field equations of the $f(R,\mathcal{G})$ gravity that are proposed in \cite{Example GaussBonnet second model f(R G)+Lm} and adopted in existing papers related to generic dependence on $\mathcal{G}$.
On the other hand, recall that in \emph{four} dimensions the Gauss-Bonnet invariant $\mathcal{G}$ is proportional to the Euler-Poincar\'e topological density as
\begin{equation}
\mathcal{G}=\Big(\frac{1}{2}\epsilon_{\alpha\beta\gamma\zeta}R^{\gamma\zeta\eta\xi}\Big)\cdot \Big(\frac{1}{2}\epsilon_{\eta\xi\rho\sigma} R^{\rho\sigma\alpha\beta}\Big)
={}^*R_{\alpha\beta}^{\;\;\;\;\,\eta\xi}
\,{}^*R_{\eta\xi}^{\;\;\;\;\,\alpha\beta}\;,
\end{equation}
where $\epsilon_{\alpha\beta\gamma\zeta}$ refers to the totally antisymmetric Levi-Civita (pseudo)tensor with $\epsilon_{0123}=\sqrt{-g}$. The integral $\int dx^4 \sqrt{-g}\,\mathcal{G}$ is equal to the Euler characteristic number $\chiup$ (just a constant) of the spacetime, and thus
\begin{equation}\label{GaussBonnet Bach-Lanczos identity}
\frac{\delta}{\delta g^{\mu\nu}}\int dx^4 \sqrt{-g}\,\mathcal{G} \,\equiv\,0\;.
\end{equation}
By explicitly carrying out this variational derivative, one could find the following Bach-Lanczos identity \cite{Euler density}:
\begin{equation}\label{GaussBonnet Bach-Lanczos identity II}
2 RR_{\mu\nu}-4 R_\mu^{\;\;\,\alpha}R_{\alpha\nu}-4 R_{\alpha\mu\beta\nu}R^{\alpha\beta}
+2R_{\mu\alpha\beta\gamma}R_{\nu}^{\;\;\,\alpha\beta\gamma}\equiv \frac{1}{2}\mathcal{G}\,g_{\mu\nu},
\end{equation}
with which the standard field equations (\ref{GB Field Eqn}) and (\ref{GB Field Eqn II})  of the $f(R,\mathcal{G})$ gravity can be simplified into
\begin{equation}\label{GB Field Eqn new I}
\begin{split}
R_{\mu\nu}-\frac{1}{2}R g_{\mu\nu}\;=\;&8\pi \frac{G}{f_R}\, T_{\mu\nu}^{(m)}
+ \frac{1}{f_R}\Bigg(\frac{1}{2}\big(f-f_{\mathcal{G}}\mathcal{G}- f_R R\big)g_{\mu\nu}
+\big(\nabla_\mu\!\nabla_\nu-g_{\mu\nu}\Box\big)f_R\\
&+2R\,\big(\nabla_\mu\!\nabla_\nu-g_{\mu\nu}\Box\big)f_{\mathcal{G}}
+4R_{\mu\nu}\Box f_{\mathcal{G}}+\mathcal{H}_{\mu\nu}^{\text{(GB)}}\Bigg)\,,
\end{split}
\end{equation}
where
\begin{equation}\label{GB Field Eqn new II}
\mathcal{H}_{\mu\nu}^{\text{(GB)}} \,\coloneqq\,
-4R_{\mu}^{\;\;\,\alpha}\nabla_\alpha\!\nabla_{\nu}f_{\mathcal{G}}
-4R_{\nu}^{\;\;\,\alpha}\nabla_\alpha\!\nabla_{\mu} f_{\mathcal{G}}\\
+4g_{\mu\nu} \!\cdot\! R^{\alpha\beta}\nabla_\alpha\!\nabla_\beta f_{\mathcal{G}}-4\,R_{\alpha\mu \beta\nu} \nabla^\beta \nabla^\alpha  f_{\mathcal{G}}\,.
\end{equation}
This way, the effective gravitational coupling strength is recognized to be
\begin{equation}
G_{\text{eff}}\,=\,\frac{G}{f_R}\,,
\end{equation}
as opposed to the $G_{\text{eff}}=G/(f_R+2Rf_{\mathcal{G}})$ in Eq.(\ref{GB Geff}); this is because the $2f_{\mathcal{G}} RR_{\mu\nu}$ term directly joining Eq.(\ref{GB Field Eqn}) is now absorbed by the $\frac{1}{2}f_{\mathcal{G}}\mathcal{G}\,g_{\mu\nu}$ term in Eq.(\ref{GB Field Eqn new I}) due to the Bach-Lanczos identity and thus no longer shows up in Eq.(\ref{GB Field Eqn new I}). The SEM tensor from modified-gravity effects  becomes
\begin{equation}\label{GB new TMG}
T_{\mu\nu}^{\text{(MG)}}=
\frac{1}{8\pi G}\,\Bigg(\frac{1}{2}\big(f-f_{\mathcal{G}}\mathcal{G}- f_R R\big)g_{\mu\nu}
+\,\big(\nabla_\mu\!\nabla_\nu-g_{\mu\nu}\Box\big)f_R
+\,2R\,\big(\nabla_\mu\!\nabla_\nu-g_{\mu\nu}\Box\big)f_{\mathcal{G}}+4R_{\mu\nu}\Box f_{\mathcal{G}}+\mathcal{H}_{\mu\nu}^{\text{(GB)}} \Bigg)\,,
\end{equation}
which with the FRW metric  Eq.(\ref{FRW metric I}) gives rise to
\begin{equation}
\rho_{\text{(MG)}}\,=\,\frac{1}{8\pi G}\,\Bigg( \frac{1}{2}\,(f_R R+f_{\mathcal{G}}\mathcal{G})-\frac{1}{2}\,f-3H\dot{f}_R
-6RH\dot{f}_{\mathcal{G}}
+4R_t^{\;\;t}(\ddot{f}_{\mathcal{G}}+3H\dot{f}_{\mathcal{G}})-\mathcal{H}_{t\,\text{(GB)}}^{\;\;t}
 \Bigg)
\end{equation}
and
\begin{equation}
P_{\text{(MG)}}\,=\,
\frac{1}{8\pi G}\,\Bigg(\frac{1}{2}\,f- \frac{1}{2}\,(f_R R+f_{\mathcal{G}}\mathcal{G})
+\ddot{f}_R+2H\dot{f}_R
+2R(\ddot{f}_{\mathcal{G}}+2H\dot{f}_{\mathcal{G}})
-4R_r^{\;\;r}(\ddot{f}_{\mathcal{G}}+3H\dot{f}_{\mathcal{G}})+\mathcal{H}_{r\,\text{(GB)}}^{\;\;r}
\Bigg)\,.
\end{equation}
Since the $G_{\text{eff}}=G/f_R$ coincides with that of $f(R)$ gravity,  the Hawking or Misner-Sharp geometric mass $M$, the nonequilibrium energy dissipation $\mathcal {E}$, the horizon entropy $S$ and the  entropy production element $d_pS$ are all the same with those of $f(R)$ gravity, as derived before in Eqs.(\ref{fR mass}), (\ref{fR E S}) and (\ref{fR dpS}) in Sec.~\ref{fR gravity}, respectively. Then the thermodynamical approach of Sec.~\ref{Inside the cosmological apparent horizon} and Sec.~\ref{On the cosmological apparent horizon} yields
\begin{equation}\label{GB Friedmann new 1}
H^2+\frac{k}{a^2}=\frac{8\pi}{3}\frac{G}{f_R}\rho_m + \frac{1}{3 f_R}\Bigg( \frac{1}{2}\big(f_R R+f_{\mathcal{G}}\mathcal{G}\big)-\frac{1}{2}f
-3H\,\big(\dot{f}_R+2R\dot{f}_{\mathcal{G}}\big)
+4R_t^{\;\;t}\,\big(\ddot{f}_{\mathcal{G}}+3H\dot{f}_{\mathcal{G}}\big)-\mathcal{H}_{t\,\text{(GB)}}^{\;\;t}
\Bigg)
\end{equation}
and
\begin{equation}\label{GB Friedmann new 2}
\dot H-\frac{k}{a^2}=-4\pi\frac{G}{f_R} \big(\rho_m+P_m\big)-\frac{1}{2f_R}\Bigg( \ddot{f}_R
-H\dot{f}_R
+2R\ddot{f}_{\mathcal{G}}-2RH\dot{f}_{\mathcal{G}}
+\,4\big(R_t^{\;\;t}-R_r^{\;\;r}\big)\big(\ddot{f}_{\mathcal{G}}+3\mathcal{H}\dot{f}_{\mathcal{G}}\big)
-\mathcal{H}_{t\,\text{(GB)}}^{\;\;t}+\mathcal{H}_{r\,\text{(GB)}}^{\;\;r}
\Bigg)\,,
\end{equation}
which match the Friedmann equations obtained from substituting the FRW metric Eq.(\ref{FRW metric I}) into the simplified  $f(R,\mathcal{G})$ field equation (\ref{GB Field Eqn new I}).

However, these thermodynamical quantities and relations of $f(R,\mathcal{G})$ gravity differ dramatically with those in  the previous Sec.~\ref{fRG gravity}. The contrast may be seen even more evidently in the $\mathscr{L}=R+f(\mathcal{G})+16\pi G\mathscr{L}_m$ modified Gauss-Bonnet gravity \cite{GaussBonnet first model R/2k+f(G)} which is a special subclass of the $f(R,\mathcal{G})$ theory. It follows from Sec.~\ref{fRG gravity} that $G_{\text{eff}}=G/(1+2Rf_{\mathcal{G}})$ for $f(R,\mathcal{G})=R+f(\mathcal{G})$, and it is a nonequilibrium scenario with nonvanishing energy dissipation $\mathcal{E}$ and entropy production $d_pS$ on the apparent horizon. On the contrary, we have $G_{\text{eff}}=G$ in accordance with Eq.(\ref{GB Field Eqn new I}) as $f_R=1$, which corresponds to an equilibrium gravitational thermodynamics with $\mathcal{E}=0=d_pS$.

Note that the existence of the two distinct formulations for the thermodynamics of  $f(R,\mathcal{G})$ gravity does not indicate a failure of our unified formulation. Instead,
it reveals a \emph{self-inconsistency} feature of  the $f(R,\mathcal{G})$ theory itself. Although the simplified field equations (\ref{GB Field Eqn new I}) and (\ref{GB Field Eqn new II})  are equivalent to Eqs.(\ref{GB Field Eqn}) and (\ref{GB Field Eqn II}) in  Sec.~\ref{fRG gravity} via the identity Eq.(\ref{GaussBonnet Bach-Lanczos identity II}), practically they will behave differently with each other in any problems relying on the input of the effective coupling  strength $G_{\text{eff}}$. Moreover, we also expect this self-inconsistency of  $f(R,\mathcal{G})$ gravity to arise in other problems such as the  black-hole thermodynamics.


\subsection{Dynamical Chern-Simons gravity}\label{Dynamical Chern-Simons gravity}

So far we have applied our unified formulation to the $f(R)$, generalized Brans-Dicke, scalar-tensor-chameleon, quadratic and  $f(R,\mathcal{G})$ gravity; they are all nonequilibrium theories with nontrivial $G_{\text{eff}}$ in the coefficient of $T_{\mu\nu}^{\text{(m)}}$. As a final example we will continue to consider the (dynamical) Chern-Simons modification of GR \cite{Chern-Simons}, which is a thermodynamically equilibrium theory with  $G_{\text{eff}}=G$. Its Lagrangian density reads
\begin{equation}\label{Generalized Brans-Dicke with self-interaction potential}
\mathscr{L}_{\text{CS}}= R+\frac{a\,\vartheta}{2\sqrt{-g}} {}^*\widehat{RR}-b \,\nabla_\mu\vartheta \nabla^\mu\vartheta-V(\vartheta)+16\pi G\mathscr{L}
_m ,
\end{equation}
where $\vartheta=\vartheta(x^\mu)$ is a scalar field, $\{a,b\}$ are constants, and ${}^*\widehat{RR}$ denotes the parity-violating Pontryagin invariant
\begin{equation}
{}^*\widehat{RR}=  {}^{*}R_{\alpha\beta\gamma\delta}\,R^{\alpha\beta\gamma\delta}
=\Big(\frac{1}{2}\epsilon_{\alpha\beta\mu\nu} R^{\mu\nu}_{\;\;\;\;\gamma\delta}\Big)\,R^{\alpha\beta\gamma\delta}\,.
\end{equation}
${}^*\widehat{RR}$ is proportional to the divergence of the Chern-Simons topological current $K^\mu$ \cite{Chern-Simons}:
\begin{equation}
{}^*\widehat{RR} = -2\,\partial_\mu K^\mu  \qquad \text{and} \qquad
K^\mu =2\epsilon^{\mu\alpha\beta\gamma}
\Big(\frac{1}{2}\Gamma^\xi_{\alpha\tau}\partial_\beta\Gamma^\tau_{\gamma\xi}
+\frac{1}{3} \Gamma^\xi_{\alpha\tau}\Gamma^\tau_{\beta\eta}\Gamma^\eta_{\gamma\xi}   \Big)\,,
\end{equation}
with $\epsilon^{0123}=1/\sqrt{-g}$,  hence the name Chern-Simons gravity.
Variational derivative of $\sqrt{-g}\mathscr{L}_{\text{CS}}$ with respect to the inverse metric $g^{\mu\nu}$ yields the field equation
\begin{equation}\label{CS eq}
R_{\mu\nu}-\frac{1}{2}Rg_{\mu\nu}=8\pi G T_{\mu\nu}^{(m)}-\frac{a}{\sqrt{-g}}\, C_{\mu\nu}
+b\,\Big(\nabla_\mu\vartheta \nabla_\nu \vartheta-\frac{1}{2} g_{\mu\nu}
\nabla_{\alpha}\vartheta \nabla^{\alpha}\vartheta\Big) -\frac{1}{2}V(\vartheta) g_{\mu\nu}\,,
\end{equation}
where
\begin{equation}
\mathcal{C}_{\mu\nu}\,=\,\nabla^\alpha\vartheta\cdot
\Big( \epsilon_{\alpha\beta \gamma\mu}\nabla^\gamma R_\nu^{\;\;\beta} +
\epsilon_{\alpha\beta\gamma\nu}\nabla^\gamma R_\mu^{\;\;\beta} \Big)
+\nabla^\alpha\nabla^\beta\vartheta \cdot \Big({}^*R_{\beta\mu\nu\alpha}+
{}^*R_{\beta\nu\mu\alpha}\Big) \,.
\end{equation}
Eq.(\ref{CS eq}) directly shows that the Chern-Simons gravitational coupling strength is just Newton's constant, $G_{\text{eff}}= G$, and
\begin{equation}
T_{\mu\nu}^{\text{(MG)}}=-a\, C_{\mu\nu}
+b\,\Big(\nabla_\mu\vartheta \nabla_\nu \vartheta-\frac{1}{2} g_{\mu\nu}
\nabla_{\alpha}\vartheta \nabla^{\alpha}\vartheta\Big)-\frac{1}{2}V(\vartheta) g_{\mu\nu}.
\end{equation}
With the FRW metric Eq.(\ref{Friedmann eqn 1st}), this $T_{\mu\nu}^{\text{(MG)}}$ leads to
\begin{equation}\label{CS rho P}
\rho_{\text{(MG)}}\,=\, \frac{1}{16\pi G}\,\bigg(b\,\dot{\vartheta}^2+  V(\vartheta)\bigg)\quad\text{and}\quad
P_{\text{(MG)}}\,=\,\frac{1}{ 16\pi G}\,\bigg(b\,\dot{\vartheta}^2 -V(\vartheta)\bigg) \,.
\end{equation}
Since $G_{\text{eff}}= G=\text{constant}$, we can make use of the reduced formulation in Sec.\ref{Situations of G without dissipative energy} for equilibrium situations. The geometric mass and the horizon entropy are respectively
\begin{equation}\label{CS mass}
M\,=\,\frac{\Upsilon^3}{2G}\left( H^2+\frac{k}{a^2}\right)\quad\text{with}\quad
M_{\text A}\;\hat{=}\;\frac{\Upsilon_{\text A}}{2G}
\end{equation}
and
\begin{equation}\label{CS entropy}
S=\frac{A_{\text A}}{4G}=\frac{\pi \Upsilon_{\text A}^2}{G}\,,
\end{equation}
which are the same with those of GR \cite{Cai I}. Also, there are no energy dissipation  $\mathcal{E}$ and the on-horizon entropy production $d_pS$,
\begin{equation}
\mathcal{E}=0\quad\text{and}\quad d_pS=0\,.
\end{equation}
Following the procedures in Sec.\ref{Situations of G without dissipative energy}, for the interior and the horizon  we obtain from the thermodynamical approach that
\begin{equation}\label{CS Friedmann 1}
H^2+\frac{k}{a^2}=\frac{8\pi G}{3}\rho_m
+\frac{1}{6}\bigg(b\,\dot{\vartheta}^2+  V(\vartheta)\bigg)
\end{equation}
and
\begin{equation}\label{CS Friedmann 2}
\dot H-\frac{k}{a^2}\,=-4\pi G\, \Big(\rho_m+P_m\Big)
-\frac{b }{2}\,\dot{\vartheta}^2\,.
\end{equation}
By substituting the FRW metric Eq.(\ref{FRW metric I}) into the field equation (\ref{CS eq}), we have confirmed that Eqs.(\ref{CS Friedmann 1}) and (\ref{CS Friedmann 2}) are really the Friedmann equations of the FRW universe governed by the  Chern-Simons gravity.


\section{Conclusions}

In this paper, we have developed a unified formulation to derive the Friedmann equations from (non)equilibrium thermodynamics within modified gravity with field equations of the form $R_{\mu\nu}-Rg_{\mu\nu}/2=8\pi G_{\text{eff}}  T_{\mu\nu}^{\text{(eff)}}$. We firstly made the necessary preparations by
locating the marginally inner trapped horizon $\Upsilon_{\text{A}}$ of the expanding FRW universe as the apparent horizon of relative causality, and then rewrote the continuity equation from $\nabla^\mu(G_{\text{eff}}  T_{\mu\nu}^{\text{(eff)}})=0$ to introduce the energy dissipation element  $\mathcal{E}$ which is related with the
evolution of $G_{\text{eff}}$.

With these preparations, we began to study the thermodynamics of the FRW universe. We have generalized the Hawking and Misner-Sharp geometric definitions of mass by replacing Newton's constant $G$ with $G_{\text{eff}}$,  and calculated the total derivative of $M$ in the comoving $(t,r)$ and the areal $(t,\Upsilon)$ transverse coordinates.
Also, we have supplemented Hayward's unified first law of thermodynamics into $dE=A\bm\psi+WdV+\mathcal{E}$ with the  dissipation term $\mathcal{E}$, where the work density $W$ and the heat flux covector $\bm\psi$ are computed using the effective matter content $T_{\mu\nu}^{\text{(eff)}}$. By identifying the geometric mass $M$ enveloped by a sphere of radius $\Upsilon< \Upsilon_{\text{A}}$ as the total internal energy $E$, the Friedmann equations have been derived from the thermodynamic equality $dM=dE$.

On the horizon $\Upsilon= \Upsilon_{\text{A}}$, besides the smooth limit $\Upsilon\to \Upsilon_{\text{A}}$ of $dM=dE$ from the untrapped interior $\Upsilon< \Upsilon_{\text{A}}$ to the horizon, we have employed an alternative Clausius method. By considering the heat flow during the infinitesimal time interval $dt$ for an isochoric process using the unified first law $dE\,\hat{=}\,A_{\text{A}}\bm\psi_t+\mathcal{E}_{\text{A}}$ and the generic nonequilibrium  Clausius relation $T(dS+d_pS)\,\hat{=}\,\delta Q$ respectively,
we have obtained the second Friedmann equation  $\dot H-k/{a^2}\,\hat{=}-4\pi G_{\text{eff}}\big(\rho_{\text{eff}}+P_{\text{eff}}\big)$ from the thermodynamics equality  $T(dS+d_pS)\,\hat{=}\,\delta Q\,\hat{=}-dE\,\hat{=}-(A_{\text{A}}\bm\psi_t+\mathcal{E}_{\text{A}})$, while the first  Friedmann equation $H^2+k/{a^2} \,\hat{=}\, 8\pi G_{\text{eff}} \rho_{\text{eff}}/3$ can be recovered using  the generalized continuity equation $\dot{G}_{\text{eff}} \rho_{\text{eff}}+ G_{\text{eff}} \dot{\rho}_{\text{eff}}+3G_{\text{eff}} H \big(\rho_{\text{eff}}+P_{\text{eff}}  \big)  =0$. Here we have taken the temperature ansatz $T=1/(2\pi \Upsilon_{\text{A}})$ in \cite{Cai I} and the Wald-Kodama dynamical entropy $S\,\hat{=}\,A_{\text{A}}/(4G_{\text{eff}})$ for the horizon, and the equality $T(dS+d_pS)\,\hat{=}-(A_{\text{A}}\bm\psi_t+\mathcal{E}_{\text{A}})$ has also determined the entropy production $d_pS$ which is generally nonzero unless $G_{\text{eff}}=$ constant . In the meantime, we have adjusted the thermodynamic sign convention by the consistency between the thermodynamics of the horizon and the interior.

After developing the unified formulation for generic relativistic gravity, we have extensively discussed some important problems related to the formulation. A viability test of the generalized effective mass has been proposed, which shows that the equality between the physically defined effective mass $\mathcal{M}=\rho_{\text{eff}} V=(\rho_m+\rho_{\text{(MG)}})V$ and the geometric effective mass automatically yields the Friedmann equations. Also, we have argued that for the modified-gravity theories under discussion with minimal geometry-matter coupling, the continuity equation can be further simplified due to the Noether-compatible definition of $T_{\mu\nu}^{ (m)}$. Furthermore, we have discussed the reduced situation of the unified formulation for $G_{\text{eff}}=G=\text{constant}$ with vanishing dissipation $\mathcal{E}=0$  and entropy production $d_pS=0$, which is of particular importance for typical scalarial dark-energy models and some fourth-order gravity.


Finally, we have applied our unified formulation to the $f(R)$, generalized Brans-Dicke,  scalar-tensor-chameleon, quadratic, $f(R,\mathcal{G})$ generalized Gauss-Bonnet and dynamical Chern-Simons gravity, to derive the Friedmann equations  from thermodynamics-gravity correspondence, where compact notations have been employed to simplify the thermodynamic quantities $\{\rho_{\text{(MG)}}\,,P_{\text{(MG)}}\}$. In addition, we have  verified that, the ``generalized Misner-Sharp energy'' for $f(R)$ and scalar-tensor gravity FRW cosmology in \cite{Misner-Sharp mass III} matches the pure mass $\rho_mV$ of the physical matter content in our formulation, and then continued to reconstruct the physical mass $\rho_m V$ from the spacetime geometry for generic modified gravity. We also found the self-inconsistency of $f(R,\mathcal{G})$ gravity in such problems which require to specify the $G_{\text{eff}}$.

In our prospective studies, we will apply the unified formulation developed in this paper to the generalized second law of thermodynamics for the FRW universe, and extend our formulation to more generic theories of modified gravity which allow for nonminimal curvature-matter couplings. Moreover, we will try to loosen the restriction of spherical symmetry
and look into the problem of thermodynamics-gravity correspondence in the Bianchi classes of cosmological solutions.


\section*{Acknowledgement}

The authors are grateful to Prof. Rong-Gen Cai (Beijing) for helpful discussion. This work was financially supported by the Natural Sciences and Engineering Research
Council of Canada.


\begin{thebibliography}{100}



\bibitem{Black hole mechanics}
J M Bardeen, B Carter, S W Hawking. \emph{The four laws of black hole mechanics}. Communications in Mathematical Physics (1973), \textbf{31}(2): 161-170.\\
Jacob D Bekenstein. \emph{Black holes and entropy}. Physical Review D  (1973), \textbf{7}(8): 2333-2346.

\bibitem{de Sitte thermodynamics}
G W Gibbons, S W Hawking. \emph{Cosmological event horizons, thermodynamics, and particle creation}. Physical Review D  (1977), \textbf{15}(10): 2738-2751.

\bibitem{Jacobson 1995}
Ted Jacobson. \emph{Thermodynamics of spacetime: The Einstein equation of state}.
Physical Review Letters (1995), \textbf{75}: 1260-1263. \href{http://arxiv.org/abs/gr-qc/9504004}{arXiv:gr-qc/9504004}

\bibitem{Unruh Temperature}
W G Unruh. \emph{Notes on black-hole evaporation}.  Physical Review D (1976), \textbf{14}(4): 870-892.



\bibitem{Hayward trapping horizons}
Sean A Hayward. \emph{General laws of black-hole dynamics}. Physical Review D  (1994), \textbf{49}(12): 6467-6474.
\href{http://arxiv.org/abs/gr-qc/9303006}{arXiv:gr-qc/9303006v3}\\
Ivan Booth. \emph{Black hole boundaries}. Canadian Journal of Physics, 2005, \textbf{83}(11): 1073-1099. \href{http://arxiv.org/abs/gr-qc/0508107}{arXiv:gr-qc/0508107v2}

\bibitem{Isolated Horizons Hamiltonian}
Abhay Ashtekar, Stephen Fairhurst, Badri Krishnan. \emph{Isolated horizons: Hamiltonian evolution and the first law}. Physical Review D (2000), \textbf{62}(10): 104025. \href{http://arxiv.org/abs/gr-qc/0005083}{gr-qc/0005083}


\bibitem{de Sitte inflation}
Andrei V Frolov, Lev Kofman. \emph{Inflation and de Sitter thermodynamics}.  Journal of Cosmology and Astroparticle Physics (2003), \textbf{2003}(05): 009. \href{http://arxiv.org/abs/hep-th/0212327}{arXiv:hep-th/0212327}


\bibitem{Cai I}
Rong-Gen Cai, Sang Pyo Kim. \emph{First law of thermodynamics and Friedmann Equations of Friedmann-Robertson-Walker universe.} Journal of High Energy Physics (2005),  \textbf{2005}(02): 050.
\href{http://arxiv.org/abs/hep-th/0501055v1}{arXiv:hep-th/0501055}


\bibitem{Cai II}
M Akbar, Rong-Gen Cai. \emph{Friedmann equations of FRW universe in scalar-tensor gravity, $f(R)$ gravity and first law of thermodynamics}.  Physics Letters  B (2006), \textbf{635}(1): 7-10. \href{http://arxiv.org/abs/hep-th/0602156}{arXiv:hep-th/0602156}



\bibitem{Eling Nonequilibrium Thermodynamics}
Christopher Eling, Raf Guedens, and Ted Jacobson. \emph{Nonequilibrium thermodynamics of spacetime}. Physical Review Letters (2006), \textbf{96}(12): 121301. \href{http://arxiv.org/abs/gr-qc/0602001}{arXiv:gr-qc/0602001}




\bibitem{Cai III-2}
Rong-Gen Cai, Li-Ming Cao. \emph{Unified first law and the thermodynamics of the apparent horizon in the FRW universe.}
Physical Review D (2007), \textbf{75}(6): 064008.
\href{http://arxiv.org/abs/gr-qc/0611071v2}{arXiv:gr-qc/0611071}



\bibitem{Misner-Sharp mass}
Charles W Misner, David H Sharp. \emph{Relativistic equations for adiabatic, spherically symmetric gravitational collapse}. Physical Review (1964),  \textbf{136}(2B): B571-576.\\
Sean A Hayward. \emph{Gravitational energy in spherical symmetry}. Physical Review D  (1996), \textbf{53}(4): 1938-1949. \href{http://arxiv.org/abs/gr-qc/9408002}{arXiv:gr-qc/9408002}




\bibitem{Misner-Sharp mass III}
Rong-Gen Cai, Li-Ming Cao, Ya-Peng Hu, Nobuyoshi Ohta. \emph{Generalized Misner-Sharp energy in $f(R)$ gravity}. Physical Review D (2009), \textbf{80}(10): 104016.  \href{http://arxiv.org/abs/0910.2387v2}{arXiv:0910.2387 [hep-th]}


\bibitem{mass-like function} 
Yungui Gong, Anzhong Wang. \emph{Friedmann equations and thermodynamics of apparent horizons}. Physical Review Letters (2007), \textbf{99}: 211301. \href{http://arxiv.org/abs/0704.0793v2}{arXiv:0704.0793 [hep-th]}



\bibitem{Cai V}
M Akbar, Rong-Gen Cai. \emph{Thermodynamic behavior of Friedmann equations at apparent horizon of FRW universe}. Physical Review D (2007), \textbf{75}(08): 084003. \href{http://arxiv.org/abs/hep-th/0609128}{arXiv:hep-th/0609128}




\bibitem{Cai IV}
M Akbar, Rong-Gen Cai. \emph{Thermodynamic behavior of field equations for $f(R)$ gravity}. Physics Letters B (2007), \textbf{648}(2-3): 243-248,. \href{http://arxiv.org/abs/gr-qc/0612089}{arXiv:gr-qc/0612089}



\bibitem{Inverse brane world}
Rong-Gen Cai,  Li-Ming Cao. \emph{Thermodynamics of apparent horizon in brane world scenario}.  Nuclear Physics B (2007), \textbf{785}(1-2): 135-148. \href{http://arxiv.org/abs/hep-th/0612144}{arXiv:hep-th/0612144}\\
Ahmad Sheykhi, Bin Wang, Rong-Gen Cai. \emph{Thermodynamical properties of apparent horizon in warped DGP braneworld}. Nuclear Physics B (2007), \textbf{779}(1-2): 1-12. \href{http://arxiv.org/abs/hep-th/0701198}{arXiv:hep-th/0701198}\\
Ahmad Sheykhi, Bin Wang, Rong-Gen Cai. \emph{Deep connection between thermodynamics and gravity in Gauss-Bonnet braneworlds}. Physical Review D (2007), \textbf{76}(02): 023515. \href{http://arxiv.org/abs/hep-th/0701261}{arXiv:hep-th/0701261}



\bibitem{scalar f R phi phi2}
Kazuharu Bamba, Chao-Qiang Geng, Shinji Tsujikawa. \emph{Equilibrium thermodynamics in modified gravitational theories}. Physics Letters B (2010), \textbf{688}(1): 101-109. \href{http://arxiv.org/abs/0909.2159v3}{arXiv:0909.2159 [gr-qc]}



\bibitem{Clausius modified gravity}
Kazuharu Bamba, Chao-Qiang Geng, Shin'ichi Nojiri, Sergei D Odintsov. \emph{Equivalence of modified gravity equation to the Clausius relation}. Europhysics Letters (2010), \textbf{89}(5): 50003. \href{http://arxiv.org/abs/0909.4397}{arXiv:0909.4397 [hep-th]}



\bibitem{Dark Energy II}
Shin'ichi Nojiri, Sergei D Odintsov. \emph{Introduction to modified gravity and gravitational alternative for dark energy}. International Journal of Geometric Methods in Modern Physics (2007),  \textbf{4}(1): 115-145. \href{http://arxiv.org/abs/hep-th/0601213v5}{arXiv:hep-th/0601213}\\
Shin'ichi Nojiri, Sergei D Odintsov. \emph{Unified cosmic history in modified gravity: from $F(R)$ theory to Lorentz non-invariant models}. Physics Report (2011), \textbf{505}(2-4): 59-144. \href{http://arxiv.org/abs/1011.0544}{arXiv:1011.0544 [gr-qc]}\\
Kazuharu Bamba, Salvatore Capozziello, Shin'ichi Nojiri, Sergei D. Odintsov. \emph{Dark energy cosmology: the equivalent description via different theoretical models and cosmography tests}. Astrophysics and Space Science (2012), \textbf{342}(1): 155-228. \href{http://arxiv.org/abs/1205.3421v3}{arXiv:1205.3421 [gr-qc]}

\bibitem{AA Tian-Booth Paper}
David W Tian, Ivan Booth. \emph{Lessons from $f(R,R_c^2,R_m^2,\mathscr{L}_m)$ gravity:
Smooth Gauss-Bonnet limit, energy-momentum conservation, and nonminimal coupling}.
Physical Review D (2014), \textbf{90}(2): 024059.  \href{http://arxiv.org/abs/1404.7823}{arXiv:1404.7823 [gr-qc]}


\bibitem{Example fR}
Thomas P Sotiriou, Valerio Faraoni. \emph{$f(R)$ theories of gravity}. Review of Modiew Physics (2010), \textbf{82}, 451-497. \href{http://arxiv.org/abs/0805.1726}{arXiv:0805.1726 [gr-qc]}\\
Antonio De Felice, Shinji Tsujikawa. \emph{$f(R)$ theories}. Living Review on Relativity (2010), \textbf{13}: 3.  \href{http://arxiv.org/abs/1002.4928v2}{arXiv:1002.4928 [gr-qc]}




\bibitem{Example GaussBonnet second model f(R G)+Lm}
Guido Cognola, Emilio Elizalde, Shin'ichi Nojiri, Sergei D Odintsov, Sergio Zerbini. \emph{Dark energy in modified Gauss-Bonnet gravity:
late-time acceleration and the hierarchy problem}. Physical Review D (2006), \textbf{73}: 084007.  \href{http://arxiv.org/abs/hep-th/0601008v2}{arXiv:hep-th/0601008}



\bibitem{Example Carroll R+ f(R Rc2 Rm2)+2kLm}
Sean M Carroll, Antonio De Felice, Vikram Duvvuri, Damien A Easson, Mark Trodden, Michael S Turner. \emph{The cosmology of generalized modified gravity models}. Physical Review D (2005), \textbf{71}: 063513. \href{http://arxiv.org/abs/astro-ph/0410031}{arXiv:astro-ph/0410031}


\bibitem{Example Quardratic gravity second paper}

K S Stelle. \emph{Classical gravity with higher derivatives}. General Relativity and Gravitation (1978), \textbf{9}(4): 353-371.


\bibitem{Brans Dicke}
C Brans, R H Dicke. \emph{Mach's principle and a relativistic theory of gravitation}. Physical Review (1961), \textbf{124}(3): 925-935.



\bibitem{scalar tensor chameleeon}
A Abdolmaleki, T Najafi, K Karami. \emph{Generalized second law of thermodynamics in scalar-tensor gravity}. Physical Review D (2014), \textbf{89}(10): 104041. \href{http://arxiv.org/abs/arXiv:1401.7549}{arXiv:1401.7549 [gr-qc]}


\bibitem{Dark Energy}
Edmund J Copeland, M Sami, Shinji Tsujikawa. \emph{Dynamics of dark energy}. International Journal of Modern Physics D (2006), \textbf{15}(11): 1753-1936. \href{http://arxiv.org/abs/hep-th/0603057v3}{arXiv:hep-th/0603057}



\bibitem{Nonminimal coupling 0}
Shin'ichi Nojiri, Sergei D Odintsov. \emph{Gravity assisted dark energy dominance and cosmic acceleration}. Physics Letters B (2004), \textbf{599}(3-4): 137-142. \href{http://arxiv.org/abs/astro-ph/0403622}{arXiv:astro-ph/0403622}\\
Gianluca Allemandi, Andrzej Borowiec, Mauro Francaviglia, Sergei D Odintsov. \emph{Dark energy dominance and cosmic acceleration in first order formalism}. Physical Review D (2005), \textbf{72}(06): 063505. \href{http://arxiv.org/abs/grqc/0504057}{arXiv:gr-qc/0504057} \\
Orfeu Bertolami, Christian G Boehmer, Tiberiu Harko, Francisco S N Lobo. \emph{Extra force in $f(R)$ modified theories of gravity}.  Physical Review D (2007), \textbf{75}: 104016. \href{arxiv.org/abs/0704.1733}{arXiv:0704.1733 [gr-qc]}


\bibitem{Nonminimal coupling}
Tiberiu Harko, Francisco S N Lobo. \emph{Generalized curvature-matter couplings in modified gravity}. Galaxies (2014), \textbf{2}(3): 410-465. \href{http://arxiv.org/abs/arXiv:1407.2013}{arXiv:1407.2013}



\bibitem{scalar tensor chameleeon II}
Justin Khoury, Amanda Weltman. \emph{Chameleon cosmology}.  Physical Review D (2004), \textbf{69}(04): 044026.
\href{http://arxiv.org/abs/astro-ph/0309411v2}{arXiv:astro-ph/0309411}




\bibitem{Hawking Ellis}
Stephen W Hawking, G F R Ellis. \emph{The Large Scale Structure of Space-Time}.
Cambridge University Press, 1973.



\bibitem{Event horizon}
Nairwita Mazumder, Subenoy Chakraborty. \emph{Does the validity of the first law of thermodynamics imply that the generalized second law of thermodynamics of the universe is bounded by the event horizon?}  Classical and Quantum Gravity (2009), \textbf{26}(19): 195016.\\
K Karami, S Ghaffari. \emph{The generalized second law of thermodynamics for the interacting dark energy in a non-flat FRW universe enclosed by the apparent and event horizons}.  Physics Letters B (2010), \textbf{685}(2-3):  115-119. \href{http://arxiv.org/abs/arXiv:0912.0363}{arXiv:0912.0363 [gr-qc]}



\bibitem{Apparent not Event horizon 0}
Bin Wang, Yungui Gong, Elcio Abdalla. \emph{Thermodynamics of an accelerated expanding universe}. Physical Review D (2006), \textbf{74}(08): 083520. \href{http://arxiv.org/abs/gr-qc/0511051}{arXiv:gr-qc/0511051}


\bibitem{Apparent not Event horizon}
Petr Hajicek. \emph{Origin of Hawking radiation}. Physical Review D (1987), \textbf{36}(4): 1065-1079.



\bibitem{BH Mass Brans Dicke}
Nobuyuki Sakai, John D Barrow. \emph{Cosmological evolution of black holes in Brans-Dicke gravity}. Classical and Quantum Gravity (2001), \textbf{18}(22):  4717¨C4723. \href{http://arxiv.org/abs/gr-qc/0102024}{arXiv:gr-qc/0102024}



\bibitem{Hayward Unified first law}
Sean A Hayward. \emph{Unified first law of black-hole dynamics and relativistic thermodynamics}. Classical and Quantum Gravity (1998), \textbf{15}(10):  3147-3162. \href{http://arxiv.org/abs/gr-qc/9710089}{arXiv:gr-qc/9710089}


\bibitem{Hayward Dynamic black-hole entropy}
Sean A Haywarda, Shinji Mukohyama, M C Ashworth. \emph{Dynamic black-hole entropy}. Physics Letters A (1999), \textbf{256}(5-6): 347-350. \href{http://arxiv.org/abs/gr-qc/9810006}{arXiv:gr-qc/9810006}




\bibitem{Hawking mass}
Stephen W Hawking . \emph{Gravitational radiation in an expanding universe}.  Journal of Mathematical Physics  (1968), \textbf{9}(4): 598-604.


\bibitem{Temperature tuneling}
Rong-Gen Cai, Li-Ming Cao, Ya-Peng Hu. \emph{Hawking radiation of apparent horizon in a FRW universe}. Classical and Quantum Gravity (2009), \textbf{26}(12): 155018. \href{http://arxiv.org/abs/0809.1554}{arXiv:0809.1554 [hep-th]}



\bibitem{Wald entropy}
Robert M Wald. \emph{Black hole entropy is the Noether charge}. Physical Review D (1993), \textbf{48}(8): R3427-R3431.
\href{http://arxiv.org/abs/gr-qc/9307038}{arXiv:gr-qc/9307038}


\bibitem{Wald entropy II}
Ted Jacobson, Gungwon Kang, Robert C Myers. \emph{On black hole entropy}.  Physical Review D (1994), \textbf{49}(12):  6587-6598.
\href{http://arxiv.org/abs/gr-qc/9312023}{arXiv:gr-qc/9312023}\\
Vivek Iyer, Robert M Wald. \emph{Some properties of the Noether charge and a proposal for dynamical black hole entropy}. Physical Review D (1994),  \textbf{50}(2): 846-864. \href{http://arxiv.org/abs/gr-qc/9403028}{arXiv:gr-qc/9403028}




\bibitem{Dynamical surface gravity}
Yungui Gong, Bin Wang, Anzhong Wang. \emph{Thermodynamical properties of the Universe with dark energy}. Journal of Cosmology and Astroparticle Physics (2007), \textbf{2007}(01): 024. \href{http://arxiv.org/abs/gr-qc/0610151}{arXiv:gr-qc/0610151}



\bibitem{Dirac large numbers hypothesis}
P A M Dirac. \emph{Cosmological models and the large numbers hypothesis}. Proceedings of the Royal Society A (1974), \textbf{338}(1615):  439-446.




\bibitem{Chemistry of AdS black holes}
David Kastor, Sourya Ray, Jennie Traschen. \emph{Enthalpy and the mechanics of AdS black holes}. Classical and Quantum Gravity (2009), \textbf{26}(19): 195011. \href{http://arxiv.org/abs/0904.2765}{arXiv:0904.2765 [hep-th]}\\
M Cvetic, G W Gibbons, D Kubiznak, C N Pope. \emph{Black hole enthalpy and an entropy inequality for the thermodynamic volume}. Physical Review D (2011), \textbf{84}(02): 024037. \href{http://arxiv.org/abs/1012.2888}{arXiv:1012.2888 [hep-th]}\\
David Kubiznak, Robert B Mann. \emph{$P-V$ criticality of charged AdS black holes}. Journal of High Energy Physics (2012), \textbf{1207}: 033. \href{http://arxiv.org/abs/1205.0559}{arXiv:1205.0559 [hep-th]}




\bibitem{Nonminimal coupling fRT}
M Sharif, M Zubair. \emph{Thermodynamics in $f(R,T)$ theory of gravity}. Journal of Cosmology and Astroparticle Physics (2012), \textbf{2012}(03): 028. \href{http://arxiv.org/abs/arXiv:1204.0848}{arXiv:1204.0848 [gr-qc]}



\bibitem{Nonminimal coupling fRTmunu}
M Sharif, M Zubair. \emph{Study of thermodynamic laws in $f(R,T,R_{\mu\nu}T^{\mu\nu})$ gravity}. Journal of Cosmology and Astroparticle Physics (2013), \textbf{2013}(11): 042.



\bibitem{Nonminimal coupling I}
Tiberiu Harko. \emph{Thermodynamic interpretation of the generalized gravity models with geometry-matter coupling}. Physical Review D (2014), \textbf{90}(04): 044067. \href{http://arxiv.org/abs/1408.3465v2}{arXiv:1408.3465 [gr-qc]}



\bibitem{Euler density}
Bryce S DeWitt. \emph{Dynamical Theory of Groups and Fields}. Chapter 16, \emph{Specific Lagrangians}. Gordon and Breach, Science Publishers, 1965. \\
David Lovelock, Hanno Rund.
\emph{Tensors, Differential Forms, and Variational Principles}. New York: Dover, 1989.



\bibitem{GaussBonnet first model R/2k+f(G)}
Shin'ichi Nojiri, Sergei D Odintsov. \emph{Modified Gauss-Bonnet theory as gravitational alternative for dark energy}. Physics Letters B (2005), \textbf{631}(1-2): 1-6. \href{http://arxiv.org/abs/hep-th/0508049v2}{arXiv:hep-th/0508049}



\bibitem{Chern-Simons}
R Jackiw, S Y Pi. \emph{Chern-Simons modification of general relativity}. Physical Review D (2003), \textbf{68}(10): 104012.
\href{http://arxiv.org/abs/gr-qc/0308071v2}{arXiv:gr-qc/0308071}\\
Stephon Alexander, Nicol\'as Yunes. \emph{Chern-Simons modified general relativity}. Physics Reports (2009),  \textbf{480}(1-2): 1-55. \href{http://arxiv.org/abs/0907.2562}{arXiv:0907.2562 [hep-th]}



\end{thebibliography}
\end{document}